\documentclass[a4paper,11pt]{article}
\pdfoutput=1 

\usepackage{jheppub} 
\usepackage{mathrsfs,graphicx,rotating,amsmath,amsfonts,mathtools,booktabs,amssymb,wasysym,caption}
\usepackage{hyperref}
\usepackage{slashed}
\usepackage{natbib}
\usepackage[table,xcdraw,dvipsnames]{xcolor}
\usepackage{graphicx}
\usepackage{bbold}
\usepackage[utf8x]{inputenc}
\usepackage[english]{babel}
\usepackage{multirow,multicol}
\usepackage{epstopdf}
\usepackage{bbm}
\usepackage{appendix}
\usepackage{libertine}
\usepackage{braket}
\usepackage{wasysym}
\usepackage{empheq}
\usepackage{cancel}
\usepackage{enumitem}
\usepackage{mathrsfs}
\usepackage{lmodern}
\usepackage{tabularx}
\usepackage{multicol}
\usepackage{color}
\usepackage{mathtools}
\usepackage{framed}
\usepackage[most]{tcolorbox}

\definecolor{Gray}{gray}{0.9}
\definecolor{LightCyan}{rgb}{0.88,1,1}
\definecolor{darkred}{rgb}{0.7,0.3,0.3}
\definecolor{darkgreen}{rgb}{0.2,0.7,0.3}
\definecolor{greyish}{rgb}{.90,.90,.90}
\definecolor{greyish2}{rgb}{.96,.96,.96}
\definecolor{darkblue2}{rgb}{0.3,0.4,0.9}
\colorlet{shadecolor}{greyish2}

\DeclareMathOperator{\e}{e}

\newcommand{\D}{{\rm d}}

\title{\boldmath Black hole perturbations of massive and partially massless spin-2 fields in (anti) de Sitter spacetime}

\author[a]{Rachel A. Rosen,}
\author[a]{Luca Santoni}

\affiliation[a]{Department of Physics, Center for Theoretical Physics, Columbia University,
538 West 120th Street, New York, New York 10027, USA}

\emailAdd{rar2172@columbia.edu}
\emailAdd{luca.santoni@columbia.edu}

\abstract{We provide a systematic and comprehensive derivation of the linearized dynamics of massive and partially massless spin-2 particles in a Schwarzschild (anti) de Sitter black hole background, in four and higher spacetime dimensions. In particular, we show how to obtain the quadratic actions  for the propagating modes and recast the resulting equations of motion in a Schr\"odinger-like form.
In the case of partially massless fields in Schwarzschild de Sitter spacetime, we study the isospectrality between modes of different parity. In particular, we prove isospectrality analytically for modes with multipole number $L=1$ in four spacetime dimensions, providing the explicit form of the underlying symmetry. 
We show that  isospectrality between partially massless modes of different parity is broken in higher-dimensional Schwarzschild de Sitter spacetimes.}

\begin{document}
\maketitle
\tableofcontents

\vspace{1cm}

\section{Introduction}

When two black holes collide, the ringdown phase of the merger can be well-described by linear perturbations around a black hole background.  These modes, known as quasi-normal modes, dictate how the system returns to equilibrium after being perturbed.  The direct detection of gravitational waves produced by colliding black hole binaries by LIGO has stimulated great interest in better understanding the spectrum and properties of these quasi-normal modes around Schwarzschild black holes.  

The dynamics of the quasi-normal modes (QNMs) for linearized gravity (i.e., a massless spin-2 field) around a Schwarzschild black hole are governed by the Regge-Wheeler~\cite{Regge:1957td} and Zerilli~\cite{Zerilli:1971wd} equations. Remarkably, the two sets of even and odd QNMs exhibit the same spectrum. This property, often referred to as \textit{isospectrality},  was proven by Chandrasekhar in 1975 \cite{1975RSPSA.343..289C,10.2307/78902,Chandrasekhar:1985kt}.  Its physical origin however is still somewhat mysterious. In general, isospectrality appears to require massless perturbations, four spacetime dimensions and flat or de Sitter asymptotics.   The generalization of the perturbation equations for a massless spin-$2$ field in a Schwarzschild spacetime to arbitrary dimensions was discussed in~\cite{Kodama:2000fa,Kodama:2003jz} (see also  \cite{Hui:2020xxx} for a review). In~\cite{Konoplya:2003dd}  the quasi-normal modes were explicitly computed away from $D=4$ and the breaking of isospectrality was explicitly shown.  Isospectrality is also generically broken for massive fields in four and higher dimensional spacetimes \cite{Rosa:2011my,Brito:2013wya}.  Furthermore, the breaking of isospectrality in AdS is presented in \cite{Cardoso:2001bb,Berti:2009kk}.  A comprehensive understanding of when perturbations of novel particles should be isospectral remains elusive.

Curiously, de Sitter spacetime allows for exotic irreducible representations that have properties of both massive and massless particles.   These are known as partially massless particles \cite{Deser:1983tm,Deser:1983mm,Higuchi:1986py,Brink:2000ag,Deser:2001pe,Deser:2001us, Deser:2001wx,Deser:2001xr,Zinoviev:2001dt,Garidi:2003ys,Skvortsov:2006at, deRham:2013wv,Bernard:2017tcg}.  An analysis of massive and partially massless (PM) spin-$2$ fields on a Schwarzschild de Sitter spacetime in four dimensions has been performed in \cite{Brito:2013wya} and \cite{Brito:2013yxa}, respectively.  In particular, in \cite{Brito:2013yxa} numerical evidence was given that partially massless modes are isospectral in Schwarzschild de Sitter background.
In this work we confirm and extend these results, systematically generalizing to arbitrary spacetime dimensions and analytically proving the isospectrality of the partially massless spin-2 modes with multipole number $L=1$ in four dimensions. We also provide a consistency check of isospectrality in the large-$L$ limit.

Our set-up is the following: we consider perturbations of both massive and partially massless spin-2 particles in a Schwarzschild (anti) de Sitter background which we will denote by S(A)dS.  The physical motivation for this could be interpreted in two ways.  First, we might accept that we live in a universe in which gravity is mediated by a massless spin-2 particle and thus contains usual Schwarzschild black holes.  We could then consider the perturbations of additional massive or partially massless spin-2 particles on this background.  Second, more speculatively, we could imagine a universe in which the gravity itself is mediated by a massive spin-2 or a partially massless spin-2 particle.  Black hole solutions for massive gravitons are still poorly understood (see, e.g., \cite{Rosen:2017dvn}).  However, given current observational constraints as well as the existence of a Vainshtein mechanism in ghost-free massive gravity, it is not unreasonable to assume that astrophysical black holes in massive gravity should look perturbatively close to S(A)dS solutions.  Thus, we might imagine our setup reflecting a massive spin-2 perturbation on a massive gravity black hole.  Similarly, as there are no known theories of a single partially massless spin-2 particle with self-interactions, black hole solutions are also not known.\footnote{Monopole solutions for the free theory were found in \cite{Hinterbichler:2015nua}.}  However, we might optimistically imagine a scenario in which black hole solutions of this new theory of gravity look perturbatively close to S(A)dS in the physical regime.\\


Here is a summary of our main results.
\begin{itemize}
\item We provide a systematic and comprehensive  derivation of the linearized dynamics  of   massive and partially massless spin-2 particles in a S(A)dS black hole background, in four and higher spacetime dimensions. In particular, we show how to obtain the quadratic actions (App.~\ref{app:quadactions}) for the propagating modes and recast the resulting equations of motion in a Schr\"odinger-like form (Secs.~\ref{sec:massive4D} and \ref{sec:massiveD}). 


\item For generic values of the cosmological constant $\Lambda$, we prove that partially massless modes  of different parity with multipole number $L=1$ are isospectral in $4$-dimensional SdS spacetimes and we check that  this remains true in  the  high-multipole limit $L\gg1$ (Sec.~\ref{sec:isoPM}). This extends the well-known isospectrality in general relativity to partially massless spin-$2$ fields and confirms the numerical findings of \cite{Brito:2013yxa}.

\item We show (Sec.~\ref{sec:extremal}) that  isospectrality between partially massless modes of different parity is broken in higher-dimensional SdS spacetimes. This  parallels  what happens for massless spin-$2$ fields in general dimensions \cite{Konoplya:2003dd}.


\item We show that (Sec.~\ref{sec:extremal}), for a massive spin-2 particle, the linearized dynamics of perturbations on SdS backgrounds  generically suffers from a Gregory-Laflamme instability \cite{Gregory:1993vy,Babichev:2013una}. Its presence depends on the value of the spin-$2$ mass and it affects only the  monopole   ($L=0$). 
Our result generalizes  the findings of \cite{Brito:2013wya} to SdS spacetimes in arbitrary dimensions. 
\end{itemize}

\paragraph{Conventions.} Throughout this paper we will always work in units such that $c=\hbar=1$. In addition, we will often set to unity also the reduced Planck mass, $M_{\rm Pl}=1$. For the metric tensor, we will adopt the ``mostly-plus''  signature, $\eta=(-,+,+,+,\cdots)$. The letter $D$ denotes the  number of spacetime dimensions. Our convention for the decomposition in spherical harmonics and the Fourier transform in time is
$\Psi(t,r, \theta)= \sum_{L,M}\int \D \omega \e^{-i\omega t } \tilde \Psi(\omega,r,L,M)Y_{L}^M( \theta)$. In the following,  for simplicity, we will often omit the arguments on $\tilde{\Psi}$ altogether and drop  the tilde to denote the Fourier transform, relying on the context to discriminate between the different meanings.  
In some circumstances, the presentation may become technical and  involve quite lengthy expressions. For the reader's convenience, we have thus highlighted in grey boxes the main equations and results.

\section{Darboux transformations and isospectrality}
\label{sec:DT}
In gravity, quasi-normal modes (QNMs) are the characteristic frequencies that encode the information on how a compact object relaxes to its equilibrium configuration after being perturbed. In the case of Schwarzschild black holes in general relativity, they are obtained by solving the Regge-Wheeler~\cite{Regge:1957td} and Zerilli~\cite{Zerilli:1971wd} equations with the requirement that the solution reduces to a purely outgoing wave at the boundaries.\footnote{In the case of asymptotically-flat, Schwarzschild spacetimes, the boundaries are, respectively, at the black hole horizon and $r=+\infty$, where $r$ denotes the radial coordinate of the Schwarzschild metric. In the case of SdS, the latter is replaced by the cosmological horizon.} 
Remarkably, the two sets of even and odd QNMs turn out to be identical. This property, which is usually referred to as \textit{isospectrality},  was proven long ago by Chandrasekhar  \cite{1975RSPSA.343..289C,10.2307/78902,Chandrasekhar:1985kt} and follows from the fact that, in $D=4$, the Regge-Wheeler and Zerilli potentials belong to the same class of supersymmetric potentials \cite{Cooper:1994eh}.
Isospectrality generically ceases to hold for massive spin-$2$ perturbations in Schwarzschild or SdS spacetimes. However, as found numerically in \cite{Brito:2013yxa} and as we will give evidence for below in Sec.~\ref{sec:isoPM}, it is restored for partially massless spin-$2$ fields. Before getting there, it is useful and instructive to review the ingredients that are necessary for isospectrality to hold in general.

Let us start considering two distinct sectors, each one containing a single degree of freedom, whose linearized dynamics is described by a one-dimensional Schr\"odinger-like equation of the form  
\begin{equation}
\frac{\D^2\Psi_\pm}{\D r_\star^2} + W_\pm  \Psi_\pm =0 \, .
\label{GDT}
\end{equation}
The $\pm$ symbol is used to denote the two sectors, $W_+$ and $W_-$ are the two potentials (the dependence on the frequency is included in $W_\pm$), and   $r_\star\in(-\infty,+\infty)$ is the variable that will  identify below  the radial  tortoise coordinate. For the moment, we will keep  $W_\pm$ generic and review the conditions under which the equations \eqref{GDT} admit the same set of QNMs.

A sufficient condition for isospectrality to hold is the existence of a symmetry transformation mapping the equation for $\Psi_+$ into the equation for $\Psi_-$ and viceversa, \textit{and} such that it preserves the (outgoing) boundary conditions at $r_\star\rightarrow\pm\infty$.\footnote{We stress that the existence of a duality alone is not enough to guarantee that modes of different parity have the same QNM spectrum. Indeed, it is crucial that the  transformation preserves the boundary conditions. Consider the case of massless spin-$2$ perturbations around SAdS spacetimes: in this case, a mapping of the form \eqref{GDT2} still exists but it does not preserve the SAdS-boundary conditions, resulting in the breaking of isospectrality, as opposed to what happens in  Schwarzschild or SdS \cite{Cardoso:2001bb,Berti:2009kk}.} 
Let us consider the most general linear transformation relating the on-shell fields  $\Psi_+$ and $\Psi_-$. Given that the equations of motion are second order, we can write it in general as
\begin{equation}
\Psi_+ = \beta(r_\star) \partial_{r_\star}\Psi_- + F(r_\star) \Psi_- \, ,
\label{GDT2}
\end{equation}
which belongs to the class of the so-called (generalized) Darboux transformations discussed in \cite{Glampedakis:2017rar} and originally introduced by G.~Darboux in  \cite{darboux1999proposition}. 
In  \eqref{GDT2}, $\beta$ and $F$ are functions of $r_\star$ and they are assumed to asymptote a constant as $r_\star\rightarrow\pm\infty$. This  is crucial  in order for the duality to preserve the form of the solution at the boundaries.



Plugging \eqref{GDT2} into the equation for $\Psi_+$ and using the $\Psi_-$'s equation of motion, one can derive the following constraints on $\beta$ and $F$ \cite{Glampedakis:2017rar}:
\begin{subequations}
\label{GDT3}
\begin{align}
\label{GDT3a}
2 \partial_{r_\star}F + \partial_{r_\star}^2\beta + \beta(W_+-W_-)  & = 0 \, ,
\\
\partial_{r_\star}^2F - \beta^{-1}\partial_{r_\star} ( \beta^2 W_-) + F (W_+-W_-) & = 0 \, .
\end{align}
\end{subequations}
Solving for $F$ after simple manipulations, one finds the following integro-differential equation for $\beta$,
\begin{equation}
\frac{\partial_{r_\star}^3 \beta +2 (W_++W_-)\partial_{r_\star} \beta + \beta \partial_{r_\star}(W_++W_-)}{W_+-W_-}
= \int \D r_\star \, \beta (W_--W_+) \, .
\label{GDTc2}
\end{equation}
Thus, looking in general for a duality between even and odd sector amounts to solving the integro-differential equation \eqref{GDTc2} for $\beta$---or, equivalently, the fourth-order differential equation for $\beta$ obtained after taking the derivative of \eqref{GDTc2}.  Again, crucially, the duality only exists if the transformation one finds in this way preserves the (outgoing) boundary conditions.

Alternatively, Eqs.~\eqref{GDT3}  can be combined in a Riccati equation,
\begin{equation}
 F^2  + F \partial_{r_\star}\beta - \beta \partial_{r_\star}F  + \beta^2 W_- = \text{constant} \, ,
 \label{GDTc1}
\end{equation}
where the r.h.s.~denotes an integration constant that results from removing an overall derivative in $r_\star$.   Note that the quantity on the l.h.s.~of \eqref{GDTc1} is precisely the proportionality  factor between the Wronskians $\mathcal{W}_\pm$ associated with the equations \eqref{GDT}. 
Indeed, denoting with $\Psi^{(1,2)}_\pm$ any two linearly independent solutions in each sector, then, using  \eqref{GDT2}, one finds
\begin{equation}
\mathcal{W}_+
	 \equiv \Psi_+^{(1)} \partial_{r_\star} \Psi_+^{(2)} -  \Psi_+^{(2)} \partial_{r_\star} \Psi_+^{(1)}
	= \left( F^2  + F \partial_{r_\star}\beta - \beta \partial_{r_\star}F  + \beta^2 W_-  \right) \mathcal{W}_- \, .
\label{Wrks}
\end{equation}
The Riccati equation \eqref{GDTc1} implies that $\mathcal{W}_+ = \text{constant} \times \mathcal{W}_- $, which guarantees that, if the transformation preserves the boundary conditions, then the two sectors have a common set of QNMs, defined as the values  of the frequency for which the Wronskians vanish \cite{Kokkotas:1999bd,Nollert:1999ji,Szpak:2004sf}.\footnote{It is worth emphasizing that the proportionality factor between the Wronskians is constant in $r_\star$, but it can, and in general will, depend on the frequency $\omega$. The values of $\omega$ for which this constant factor vanishes are usually referred to as algebraically special modes, which we will disregard in the following. More details can be found e.g.~in \cite{Berti:2009kk}.}

Notice that the Chandrasekhar relation \cite{1975RSPSA.343..289C,10.2307/78902,Chandrasekhar:1985kt} for the massless spin-2 field belongs to the subclass  of transformations \eqref{GDT2} with $\beta\equiv 1$ \cite{Glampedakis:2017rar},
\begin{equation}
\Psi_+ = \partial_{r_\star}\Psi_- + F(r_\star) \Psi_- \, .
\label{GDT2--}
\end{equation}
In this case, \eqref{GDTc2} becomes a consistency condition for the potentials $W_\pm$:
\begin{equation}
\frac{ \partial_{r_\star}(W_++W_-)}{W_+-W_-}
= \int \D r_\star (W_--W_+) \, ,
\label{GDTc2-2}
\end{equation}
which is famously satisfied by the Regge-Wheeler and Zerilli potentials. The form of the Darboux transformation is then unambiguously fixed in terms of the potentials by
\begin{equation}
F = \frac{\partial_{r_\star}(W_++W_-)}{2(W_+-W_-)} \, .
\label{DTFsol}
\end{equation}

\section{Massive and partially massless spin-2 fields on S(A)dS spacetimes in 4D}
\label{sec:massive4D}

We start by deriving the equations governing the linear dynamics of massive spin-$2$ fields and partially massless spin-2 fields on $4$-dimensional Schwarzschild-(A)dS spacetimes.\footnote{Results for massive spin-2 fields on pure Schwarzschild backgrounds have  been discussed in \cite{Brito:2013wya}. As opposed to \cite{Brito:2013wya}, here we mainly  work  at the level of the action \eqref{Sspin-2massive}  and in the presence of a non-zero cosmological constant.}
In particular, we will show how to obtain the quadratic action for the propagating degrees of freedom and  the corresponding Schr\"odinger-like equations of motion.
In the partially massless case, we will study the property of isospectrality that was advocated in  \cite{Brito:2013yxa} on the basis of  an explicit  numerical computation of the QNMs. We will provide an analytic proof of this fact in two cases: for modes with $L=1$ and in the eikonal limit ($L\gg 1$).

Our starting point is the quadratic Fierz-Pauli action for a spin-$2$ field $h_{\mu\nu}$ of generic mass $m$ in a $4$-dimensional S(A)dS spacetime,
\begin{multline}
S=\int \D^4x\sqrt{-g}\left[ -{1\over 2}\nabla_\lambda h_{\mu\nu} \nabla^\lambda h^{\mu\nu}+\nabla_\lambda h_{\mu\nu} \nabla^\nu h^{\mu\lambda}-\nabla_\mu h\nabla_\nu h^{\mu\nu}+\frac{1}{2} \nabla_\mu h\nabla^\mu h 
\right.
\\
\left.
+\frac{R}{4}\left( h^{\mu\nu}h_{\mu\nu}-\frac{1}{2} h^2\right)
-\frac{1}{2}m^2\left( h^{\mu\nu}h_{\mu\nu}- h^2\right) 
\right] \, .
\label{Sspin-2massive4}
\end{multline}
The field $h_{\mu\nu}$ is assumed to propagate on a fixed S(A)dS background, whose metric $g_{\mu\nu}$ is given by the usual form
\begin{equation}
\D s^2 = g_{\mu\nu}\D x^\mu\D x^\nu = -f(r) \D t^2 + \frac{1}{f(r)}\D r^2+r^2 \D\Omega_{S^2}^2,
\label{eq:sphericalmetric}
\end{equation}
where
\begin{equation}
f(r)  = 1-\frac{r_s}{r}-\frac{\Lambda}{3}  r^2.
\label{eq:sadsf}
\end{equation}
 In the following, we will mostly assume the cosmological constant $\Lambda$ to be positive (SdS spacetime), but all our expressions  hold also on a SAdS spacetime.  On this background, the massive spin-2 field in 4  dimensions propagates 5 degrees of freedom, which correspond to the $\pm$ helicity-2 modes, the $\pm$ helicity-1 modes and the helicity-0 mode.
 
At a special value of the mass relative to the background de Sitter curvature $m^2 = \tfrac{2}{3} \Lambda$, the action \eqref{Sspin-2massive4} acquires a gauge symmetry of the form
\begin{equation}
\delta h_{\mu\nu} = \left( \nabla_\mu\nabla_\nu + \frac{\Lambda}{3} g_{\mu\nu} \right)\epsilon \, ,
\label{gaugetrans}
\end{equation}
with gauge parameter $\epsilon$.  This symmetry is responsible for removing the helicity-0 component from the particle's spectrum.  These particles thus propagate only four degrees of freedom and are referred to as ``partially massless" \cite{Deser:1983tm,Deser:1983mm,Higuchi:1986py,Brink:2000ag,Deser:2001pe,Deser:2001us, Deser:2001wx,Deser:2001xr,Zinoviev:2001dt,Garidi:2003ys,Skvortsov:2006at, deRham:2013wv,Bernard:2017tcg}.  They are special irreducible representations of massive spinning particles that can propagate on  Einstein spacetimes as well as some more general spacetimes \cite{Bernard:2017tcg}.  In App.~\ref{app:pmreview}, we briefly review the main features of partially massless fields in $4D$ S(A)dS spacetimes. 

The invariance of the background metric in Eq.~\eqref{Sspin-2massive4} under spatial rotations allows one to decompose the tensor field in spherical harmonics \cite{Regge:1957td}. In particular, one can distinguish between polar (even) and axial (odd) components. The fact that parity is not broken, neither explicitly not spontaneously, guarantees that  propagating degrees of freedom of different parity do not mix at the level of the linearized equations of motion.  The 10 components of $h_{\mu\nu}$ decompose into 3 odd components and 7 even components.  In the odd sector, these 3 components yield no modes for $L=0$, one mode for $L=1$ and 2 modes for $L\geq 2$.  In the even sector for the massive spin-2 particle, the 7 metric components give one mode for $L=0$, two modes for $L=1$ and three modes for $L\geq 2$.  For the partially massless particle, the number of modes in the even sector is reduced by one at each $L$.

\subsection{Odd sector}
\label{sec:massiveodd4D}

Let us start with the odd sector. The most general parametrization of axial spin-$2$ perturbations in $4$ dimensions takes on the form \cite{Regge:1957td}\footnote{For the reader's convenience, we  note that the definition of $h_2$ differs by a sign with respect to \cite{Regge:1957td,Franciolini:2018uyq}.}
\begin{equation}
h_{\mu\nu}^{\text{odd}} = \begin{pmatrix}
0 & 0 & - \frac{h_0(t,r)}{\sin\theta} \partial_\phi &  h_0(t,r)\sin\theta \partial_\theta \\
* & 0 & - \frac{h_1(t,r)}{\sin\theta} \partial_\phi & h_1(t,r)\sin\theta \partial_\theta \\
* & * & -\frac{h_2(t,r)}{\sin\theta}\left( \partial_{\theta}\partial_{\phi} - \frac{\cos\theta}{\sin\theta}\partial_\phi \right) & \frac{1}{2}h_2(t,r) \sin\theta\left( \partial_{\theta}^2 - \frac{\cos\theta}{\sin\theta}\partial_\theta - \frac{1}{\sin^2\theta} \partial_\phi^2 \right) \\
* & * & * &   h_2(t,r)\sin\theta\left( \partial_{\theta}\partial_{\phi} - \frac{\cos\theta}{\sin\theta}\partial_\phi \right)
\end{pmatrix}  Y_{L}^M(\theta,\phi) ,
\label{hPModd4D}
\end{equation}
(the asterisks denote symmetric components) where $Y_{L}^M(\theta,\phi)$ are the spherical harmonics in $D=4$, which are normalized as $\int \D\Omega_2 \, Y_{L}^M(\theta,\phi)^* Y_{L'}^{M'}(\theta,\phi) =\delta_{LL'}\delta^{MM'}$, and where $h_0$, $h_1$ and $h_2$ are pseudo-scalars.
By construction, the components of the tensor field $h_{\mu\nu}^{\text{odd}}(t,r,\theta,\phi)$  pick up a factor of $(-1)^{L+1}$ under a parity transformation, $(\theta, \phi)\rightarrow(\pi-\theta, \phi+\pi)$. This is  why they are referred to as odd, or axial, modes.

For non-zero values of the mass of the spin-$2$ field, one expects two propagating degrees of freedom in the odd sector if $L\geq2$ corresponding to the odd helicity-1 and helicity-2 modes. This means that one of the three components of the tensor \eqref{hPModd4D} corresponds to a non-dynamical variable. Plugging the field decomposition \eqref{hPModd4D} into the Fierz-Pauli action \eqref{Sspin-2massive4}, one can easily derive the equations of motion. It is not hard to find a linear combination of these equations that is algebraic in $h_0$. Using this to integrate  $h_0$ out and plugging the solution back into the equations for $h_1$ and $h_2$, one finds the following system of coupled differential equations:
\begin{subequations}
\label{PModdL24D}
\begin{tcolorbox}[colframe=white,arc=0pt,colback=greyish2] \vspace{-0.4cm}
\begin{align}
\frac{\D^2}{\D r_\star^2} Q +  \left( \omega^2 - V_Q \right) Q  & = S_Z Z \, ,
\\
\frac{\D^2}{\D r_\star^2} Z +  \left( \omega^2 - V_Z \right) Z  & = S_Q Q \, ,
\end{align}
\end{tcolorbox}
\end{subequations}
\noindent where we have defined the fields 
\begin{subequations}
\begin{align}
Q(t,r) & \equiv - f(r) h_1(t,r) \, ,
\label{Scspin2odd4D-1}
\\
Z(t,r)  & \equiv  \frac{1}{2r} h_2(t,r) \, ,
\label{Scspin2odd4D-2}
\end{align}
\label{Scspin2odd4D}
\end{subequations}
and where 
\begin{subequations}
\label{VSodd4D}
\begin{align}
V_Q &= f \left(m^2 -  \frac{2 \Lambda }{3}  - \frac{8 r_s}{r^3}  + \frac{L^2+L+4 }{r^2}  \right) \, ,
\label{VQodd4D}
\\
V_Z &= f \left(  m^2 - \frac{2 \Lambda }{3}  +  \frac{r_s}{r^3} +\frac{L^2+L-2}{r^2}  \right) \, ,
\\
S_Z &= f \frac{\left(L^2+L-2\right) (2 r-3 r_s)}{r^3} \, ,
\\
S_Q &= \frac{2f}{r^2} \, ,
\end{align}
\end{subequations}
which agree with \cite{Brito:2013wya} in the limit $\Lambda\rightarrow0$. $r_\star$ is the tortoise coordinate, defined by
\begin{equation}
\frac{\D r_\star}{\D r} = \frac{1}{f(r)} \, .
\end{equation}

Modes with $L=1$ deserve a separate discussion. Indeed, when $L=1$ there is only one propagating degree of freedom in the odd sector.  In this case, $h_2$ drops out from the field decomposition \eqref{hPModd4D} and one is left with an action that depends exclusively on $h_0$ and $h_1$. After analogous manipulations, one finds the following  Schr\"odinger-like equation:
\begin{tcolorbox}[colframe=white,arc=0pt,colback=greyish2]
\begin{equation}
\label{massiveoddL14D}
\frac{\D^2}{\D r_\star^2} Q +  \left( \omega^2 - V_Q^{(L=1)} \right) Q =0 \, ,
\end{equation}
\end{tcolorbox}
\noindent where $Q$ is defined as in \eqref{Scspin2odd4D-1} while the potential $V_Q^{(L=1)}$ can be read off from Eq.~\eqref{VQodd4D} after setting $L=1$,
\begin{equation}
V_Q^{(L=1)} = f \left(m^2 -  \frac{2 \Lambda }{3}  - \frac{8 r_s}{r^3}  + \frac{6 }{r^2}  \right) \, .
\label{VQodd4DL1}
\end{equation}
This completes the analysis of the linearized dynamics for massive spin-$2$ axial perturbations in $D=4$ (recall that there is no propagating degree of freedom with $L=0$ in the odd sector).

The equations \eqref{PModdL24D} and \eqref{massiveoddL14D} have been derived by integrating out the  non-dynamical variable $h_0$ at the level of the fields' equations of motion.
It is instructive, however,  to obtain an expression for the quadratic action of the canonically normalized propagating fields. This is discussed  in App.~\ref{app:quadactions}.\\
\\
{\bf Partially Massless:} The partially massless equations for the parity odd modes follow simply from setting $m^2=2\Lambda/3$ in Eqs.~\eqref{PModdL24D}, \eqref{Scspin2odd4D} and \eqref{VSodd4D}.  In particular, the potentials are now given by
\begin{subequations}
\label{VSodd4D-2}
\begin{align}
V_Q &= f \left(  \frac{L^2+L+4 }{r^2}  - \frac{8 r_s}{r^3} \right) \, ,
\label{VQodd4D-2}
\\
V_Z &= f \left(  \frac{r_s}{r^3} +\frac{L^2+L-2}{r^2}  \right) \, ,
\\
S_Z &= f \frac{\left(L^2+L-2\right) (2 r-3 r_s)}{r^3} \, ,
\\
S_Q &= \frac{2f}{r^2} \, ,
\end{align}
\end{subequations}
in agreement with \cite{Brito:2013yxa}.  Similarly, the potential $V_Q^{(L=1)}$ for the odd modes with $L=1$ can be inferred from \eqref{VQodd4DL1} after setting $m^2=2\Lambda/3$:
\begin{equation}
V_Q^{(L=1)} = 2f \frac{3 r-4 r_s}{r^3} \, .
\label{VoddL1}
\end{equation}

\subsection{Even sector}
\label{subsec:evenmassive4D}

In four dimensions, the parity-even components of a spin-$2$ field on a spherically symmetric background can be written in general as follows \cite{Regge:1957td},\footnote{Again, the notation for parity-even perturbations in \eqref{hPMeven4D} is chosen in such a way to mirror the one of \cite{Regge:1957td,Franciolini:2018uyq}. However, there is a difference in the definition of $G$, which differs by the subtraction of a trace. }
\begin{equation}
h_{\mu\nu}^{\text{even}} = \begin{pmatrix}
f H_0 & H_1 & \mathcal{H}_0 \partial_\theta &  \mathcal{H}_0 \partial_\phi \\
* & \frac{H_2}{f} & \mathcal{H}_1 \partial_\theta &  \mathcal{H}_1 \partial_\phi \\
* & * & r^2\left(\mathcal{K} + G \, \mathcal{W}  \right) &   r^2   G \left(\partial_\theta  \partial_\phi - \frac{\cos\theta }{\sin\theta}\partial_\phi  \right)     \\
* & * & * &   r^2  \sin^2\theta \left(\mathcal{K}   - G \, \mathcal{W}  \right) 
\end{pmatrix} Y_{L}^M(\theta,\phi) \, ,
\label{hPMeven4D}
\end{equation}
where $\mathcal{W} $ is the differential operator defined by $\mathcal{W} \equiv \frac{1}{2}( \partial_\theta^2 -  \frac{\cos\theta}{\sin\theta}\partial_\theta -\frac{1}{\sin^2\theta}\partial_\phi^2 )$.
In analogy with the odd sector, we start considering  modes with $L\geq2$. Modes with  $L=1$ and $L=0$ will be treated  separately. When $L\geq2$, a massive spin-$2$ field in $D=4$ propagates $3$ independent degrees of freedom corresponding to the even helicity-0, helicity-1 and helicity-2 modes.  Therefore, out of the $7$ components in the tensor \eqref{hPMeven4D}, only $3$ correspond to physical modes, while the others play the role of constraints or non-dynamical fields. In the following, we will review how to integrate out the non-dynamical components from the Fierz-Pauli action \eqref{Sspin-2massive} and derive the equations for the physical perturbations. The discussion follows analogous steps  to \cite{Kobayashi:2014wsa,Franciolini:2018uyq,Franciolini:2018aad}. In particular, we will find that $H_0$ plays the role of a Lagrange multiplier enforcing a constraint that allows us to integrate $\mathcal{H}_1$ out, while $H_1$ can be eliminated using its own equation of motion.

After plugging the field decomposition \eqref{hPMeven4D} into the action \eqref{Sspin-2massive}, one immediately notices that the field component $H_0$ never appears quadratically. In other words, it plays the role  of a Lagrange multiplier enforcing the following constraint condition,
\begin{multline}
4L (L+1)   \left( r  f'+f\right)\mathcal{H}_1
+8 L(L+1) r f \mathcal{H}'_1
-8 r^2 f  H_2' 
\\
+ 2  r  \left(  r^2 f'' -2 r f'- 4 f-2   L(L+1)  -2   m^2 r^2  + 2 \Lambda  r^2\right) H_2
\\
+ 8 r  \left( f  +   r f'- \frac{1}{2} L(L+1)-r^2  m^2+ r^2 \Lambda\right)\mathcal{K}
\\
+ 4r^2 \left(r f'+6  f   f\right)\mathcal{K}' 
+ 8r^3 f  \mathcal{K}''
-2r  L (L-1)  \left(2 +L(L+3) \right) G  =0 \, .
\label{constrH04D}
\end{multline}
We can redefine the field $H_2$ in such a way to cancel the terms proportional to $\mathcal{H}_1$ and $\mathcal{K}''$ \cite{Kobayashi:2014wsa,Franciolini:2018uyq,Franciolini:2018aad}. Introducing $\psi$ via the  combination
\begin{equation}
H_2(t,r) \equiv \frac{ L (L+1) }{ r} \mathcal{H}_1(t,r)  -\frac{1}{2r f}  \psi(t,r) +r \mathcal{K}'(t,r) \, ,
\label{shiftH24D}
\end{equation}
the constraint equation \eqref{constrH04D}  becomes algebraic in $\mathcal{H}_1$ and it can be easily solved 
as follows,
\begin{multline}
\mathcal{H}_1
= 
\frac{2r^{2} \left( 2 L(L+1) +2r^2  m^2 -2r^2 \Lambda - r^2 f'' -4 f\right)    }{2 L  (L+1) \left( r^2 f''+ 4 f-2  L(L+1) -2r^2 m^2+2r^2 \Lambda \right)}\mathcal{K}' 
\\
+\frac{ 2r  \left(  L(L+1)+r^2 (2 m^2 -2 \Lambda )- 2 f -2 r f' \right) }{ L  (L+1) \left( r^2 f''+ 4 f-2  L(L+1) -2r^2 m^2+2r^2 \Lambda \right)} \mathcal{K} 
\\
-\frac{2 r }{L (L+1) \left( r^2 f''+ 4 f-2  L(L+1) -2r^2 m^2+2r^2 \Lambda \right)}\psi'
\\
+\frac{   r^2 f''+2 r f'-2 \left( L(L+1)+r^2 m^2 -r^2 \Lambda \right)     }{2 f L(L+1) \left( r^2 f''+ 4 f-2  L(L+1) -2r^2 m^2+2r^2 \Lambda \right)}\psi
\\
+ \frac{  (L-1) r \left(2+L(L+3) \right)   }{(L+1) \left( r^2 f''+ 4 f-2  L(L+1) -2r^2 m^2+2r^2 \Lambda \right)} G  \, .
\label{calH14D}
\end{multline} 
In addition, we notice that in the action \eqref{Sspin-2massive} the only non-trivial quadratic term in  $H_1$   is without (temporal or radial) derivatives. This means that the equation of motion for $H_1$  is algebraic in $H_1$  itself, which therefore can be  integrated out straightforwardly as follows,
\begin{multline}
H_1 = \frac{2 r \dot{H}_2}{r^2 f''+2 r f'-L (L+1)-r^2 \left(m^2-2 \Lambda \right)}
\\
-\frac{L (L+1) \mathcal{H}_0'}{r^2 f''+2 r f'-L (L+1)-r^2 \left(m^2-2 \Lambda \right)}
+\frac{L (L+1) f' \mathcal{H}_0}{f \left(r^2 f''+2 r f'-L (L+1)-r^2 \left(m^2-2 \Lambda \right)\right)}
\\
-\frac{L (L+1) \dot{\mathcal{H}}_1}{r^2 f''+2 r f'-L (L+1)-r^2 \left(m^2-2 \Lambda \right)}
+\frac{r \left(r f'-2 f\right) \dot{\mathcal{K}}}{f \left(r^2 f''+2 r f'-L (L+1)-r^2 \left(m^2-2 \Lambda \right)\right)}
\\
-\frac{2 r^2 \dot{\mathcal{K}}' }{r^2 f''+2 r f'-L (L+1)-r^2 \left(m^2-2 \Lambda \right)} \, ,
\end{multline}
where in place of $H_2$ and $\mathcal{H}_1$ one should substitute the expressions \eqref{shiftH24D} and \eqref{calH14D}, respectively. Notice that, after plugging the solutions to the constraint equations  back into the action, higher derivative terms, such as $\dot{\psi}'^2$ and  $\dot{\mathcal{K}}'^2$, will generically appear. However, these can be eliminated via a redefinition of  $\mathcal{H}_0$, which in $D=4$ reads
\begin{equation}
\mathcal{H}_0 \equiv   \chi
-\frac{ r^2}{ L (L+1)}  \dot{\mathcal{K}}
+\frac{2 r^5 r_s^3 \dot{\psi}}{L (L+1) \left[2 (L-1)  (L+2) r_s^3 r^{4 }+2 m^2 r_s^3 r^{6}+ 6 r^3 r_s^{4 }\right]} \, .
\end{equation}
Then, the final quadratic action takes the following form:
\begin{equation}
\begin{split}
S_{\text{even},M=0} & = \sum_{L=2}^\infty \int\D r\D t \big[ A_{\psi\psi }\dot\psi^2 + B_{\psi\psi }\psi'^2  + C_{\psi\psi }\psi^2 
+ A_{\mathcal{K}\mathcal{K} }\dot{\mathcal{K}}^2 + B_{\mathcal{K}\mathcal{K} }\mathcal{K}'^2  + C_{\mathcal{K}\mathcal{K} }\mathcal{K}^2 
\\
&\quad
+ A_{GG }\dot G^2 + B_{GG }G'^2  + C_{GG }G^2 
+ B_{\chi\chi }\chi'^2  + C_{\chi\chi}\chi^2 
\\
&\quad + A_{\psi\mathcal{K} }\dot \psi \dot{\mathcal{K}}
+ A_{\psi G} \dot \psi \dot{G}
+ A_{G\mathcal{K} }\dot G \dot{\mathcal{K}}
+ B_{\psi\mathcal{K} }\psi'\mathcal{K}' 
+ B_{\psi G }\psi' G'
+ B_{G\mathcal{K} }G'\mathcal{K}' 
\\
&\quad
+ C_{\psi\mathcal{K} }\psi\mathcal{K}
+ C_{\psi G }\psi G
+ C_{G\mathcal{K} }G\mathcal{K}
+ D_{\psi\mathcal{K} }\psi'\mathcal{K}
+ D_{\psi G }\psi' G
+ D_{\mathcal{K}G }\mathcal{K}' G
\\
&\quad
+ E_{\psi\chi}\dot{\psi}\chi' + F_{\psi\chi}\dot{\psi}\chi
+ E_{\mathcal{K}\chi}\dot{\mathcal{K}}\chi' + F_{\mathcal{K}\chi}\dot{\mathcal{K}}\chi
\\
&\quad
+ E_{G\chi}\dot{G}\chi' + F_{G\chi}\dot{G}\chi \big]
 ,
\end{split}
\label{action4DMassive}
\end{equation}
where we restricted ourselves to modes with magnetic quantum number $M=0$ (the spherical symmetry guarantees that the other modes with $M\neq 0$ satisfy the same equations of motion) and where the  coefficients are the same of \eqref{LPMevenL2fin} below
with $D=4$. To extract the dynamics for the physical degrees of freedom, one additional step is required at this point. Notice that in the action \eqref{action4DMassive} the field $\chi$ does not have a kinetic term, suggesting that it corresponds to the residual Hamiltonian constraint. Indeed, after taking the variation of the action \eqref{action4DMassive} with respect to various fields  and after straightforward  manipulations, we  find the following algebraic  equation for $\chi$, containing only first derivatives,
\begin{multline}
-i \omega  \frac{18 (L+1) L r \left(\left(L^2+L-2\right) r+ m^2 r^3+3 r_s\right)^2}{\Lambda  r^3-3 r+3 r_s} \chi = 
\\
=  \Big[   -L (-L^3-2 L^2+L+2) r 
 \Big( r^3 ((6 L^2+6 L+3) m^2-\Lambda  (L^2+L-2)) 
  \\  
 +3 (L^4+2 L^3-L-2) r  +6 (L^2+L+1) r_s+3 m^2 r^5 (m^2-\Lambda )   \Big)    \Big]  
G
\\
+2 \Big[ 
M r^6 \left(6 (L^2+L+2) M-5 \Lambda  (L^2+L-2)\right)
\\
+(L^2+L-2) r^4 \left(3 (L^2+L+5) m^2-\Lambda (L^2+L-2)\right)
\\
+3 r^2 (L^4+2 L^3-3 L^2-4 L-9 m^2 r_s^2+4)-24 (L^2+L-2) m^2 r^3 r_s
\\
-6 (2 L^4+4 L^3-3 L^2-5 L+2) r r_s-27 L (L+1) r_s^2+3 m^4 r^8 (m^2-2 \Lambda )-18 m^4 r^5 r_s
  \Big]  \mathcal{K}
 \\
-3 \left[
 L^2 (4 m^2 r^3-2 r+3 r_s)+2 L^4 r+4 L^3 r+L (4 m^2 r^3-4 r+3 r_s)+m^2 r^2 (2 m^2 r^3-3 r_s)
 \right] \psi
 \\
+ \left[
-L (-L^3-2 L^2+L+2) r (-\Lambda  r^3+3 r-3 r_s) \left((L^2+L-2) r+m^2 r^3+3 r_s\right) 
\right] G'
\\
+\left[ -2 r (-2 + L + L^2 + 2 M r^2) ((-2 + L + L^2) r + m^2 r^3 + 
   3 rs) (-3 r + 3 rs + r^3 \Lambda)
\right] \mathcal{K}'
\\
- r \Big[ 2 r^3 \left(3(L^2+L+2) m^2-\Lambda  (L^2+L-2)\right)+3 (L^4+2 L^3+L^2-4) r 
\\
+3 (L^2+L+4) r_s+3 m^2 r^5 (M-2 \Lambda )-9 m^2 r^2 r_s
\Big] \psi ' \, ,
\end{multline}
where we Fourier-transformed in time. Plugging back into the other equations and redefining the fields as
\begin{subequations}
\begin{equation}
\psi(r) = r_s\frac{\left(L^2+L-2\right) r+m^2 r^3+3 r_s}{r^2} \tilde{\psi}(r) \, ,
\qquad
\mathcal{K}(r) = \frac{r_s^2}{r^2} \tilde{\mathcal{K}} (r) \, ,
\end{equation}
\begin{equation}
G(r) = \frac{1}{r} \tilde{G}(r) + \frac{2r_s^2}{L (L+1) r^2} \tilde{\psi}(r)
-\frac{2r_s^2}{L (L+1) r^2} \tilde{\mathcal{K}} (r) \, ,
\end{equation}
\end{subequations}
we find the following system of Schr\"odinger-like equations,
\begin{subequations}
\begin{tcolorbox}[colframe=white,arc=0pt,colback=greyish2] \vspace{-0.4cm}
\begin{align}
\frac{\D^2\tilde \psi}{\D r_\star^2} + \left( \omega^2 - V_\psi\right) \tilde \psi & = S_{\mathcal{K}}^{(\psi)}\tilde{\mathcal{K}} + S_{G}^{(\psi)}\tilde  G \, ,
\\
\frac{\D^2\tilde{\mathcal{K}}}{\D r_\star^2} + \left( \omega^2 - V_{\mathcal{K}}\right) \tilde{\mathcal{K}} & = S_{\psi}^{(\mathcal{K})} \tilde{\psi} + S_{G}^{(\mathcal{K})}\tilde  G \, ,
\\
\frac{\D^2\tilde{G}}{\D r_\star^2} + \left( \omega^2 - V_{G}\right) \tilde{G} & = S_{\psi}^{(G)} \tilde{\psi} + S_{\mathcal{K}}^{(G)}\tilde{\mathcal{K}} \, ,
\end{align}
\end{tcolorbox}
\label{massiveeqsD4}
\end{subequations}
\noindent where the explicit expressions for the coefficients are reported in App.~\ref{app:pmcoeffs}.
The final equations \eqref{massiveeqsD4} describe the dynamics of a massive spin-$2$ field, with azimuthal numbers $L\geq2$, in a $4$-dimensional S(A)dS spacetime.  

The equations for the modes with $L=1$ can be obtained analogously by substituting the value $L=1$ into the action \eqref{action4DMassive}. As a result, the field component $G$ drops out and one can follow similar steps to integrate out $\chi$ and obtain a system of two differential equations for the two propagating degrees of freedom. 

Similar considerations hold for the  modes with $L=0$. The equation for these monopole-type perturbations on a SdS background in $D=4$ was studied in detail in \cite{Brito:2013wya}. Thus, we will not report it here. In Sec.~\ref{sec:massiveD} below, we will generalize the result of  \cite{Brito:2013wya} to arbitrary dimensions. The $4$-dimensional case can be simply read off from Sec.~\ref{sec:massiveD} after substituting $L=0$.\\
\\
{\bf Partially Massless:} For the partially massless particle, the even sector must be treated more carefully than the odd sector.\footnote{Notice that the equations for the even modes in $D=4$ were derived in \cite{Brito:2013yxa}. However, the final expressions are not in a Schr\"odinger-like form. Even if this is an artifact of integrating out certain field components and it does not affect at all the observables, like the quasi-normal frequencies, it might be convenient to have equations of motion in a canonical form. In the present section, we will address this issue. The derivation of the quadratic action for the propagating components of the partially massless field is discussed in App.~\ref{app:quadactions}.}   
On a spherically-symmetric background, the partially massless gauge transformation \eqref{gaugetrans} acts non-trivially only on the partially massless components of even-type. Expanding $\epsilon$ in spherical harmonics, $\epsilon(t,r,\theta,\phi) = \sum_{L,M} \epsilon (t,r) Y^M_L(\theta,\phi)$, allows us to rewrite \eqref{gaugetrans} in components as follows,
\begin{subequations}
\begin{align}
\delta H_0 & = \frac{\ddot{\epsilon}}{f} -\frac{1}{2}f'\epsilon' -\frac{\Lambda}{3}\epsilon \, ,
\\
\delta H_1 & = \dot{\epsilon}' - \frac{f'}{2f}\dot{\epsilon} \, ,
\\
\delta H_2 & = f \epsilon'' + \frac{1}{2}f' \epsilon'  + \frac{\Lambda}{3}\epsilon \, ,
\\
\delta \mathcal{H}_0 & = \dot{\epsilon} \, ,
\\
\delta \mathcal{H}_1 & = \epsilon' - \frac{\epsilon}{r} \, ,
\\
\delta \mathcal{K} & =  \frac{ f}{ r} \epsilon'  +  \left(\frac{ \Lambda  }{3}  -\frac{ L(L+1)}{2 r^2}\right) \epsilon \, ,
\\
\delta G & = \frac{\epsilon}{r^2} \, .
\end{align}
\end{subequations}
The partially massless gauge symmetry \eqref{gaugetrans} thus gives us the freedom to eliminate one of the field components in  \eqref{hPMeven4D}. In particular, we shall fix $G=0$.  The procedure follows exactly the one above---see also \cite{Franciolini:2018uyq,Franciolini:2018aad}---and the final result can be read off from Eq.~\eqref{action4DMassive} upon fixing $G=0$ and $m^2=2\Lambda/3$,
\begin{equation}
\begin{split}
L & = A_{\psi\psi }\dot\psi^2 + B_{\psi\psi }\psi'^2  + C_{\psi\psi }\psi^2 
+ A_{\mathcal{K}\mathcal{K} }\dot{\mathcal{K}}^2 + B_{\mathcal{K}\mathcal{K} }\mathcal{K}'^2  + C_{\mathcal{K}\mathcal{K} }\mathcal{K}^2 
+ B_{\chi\chi }\chi'^2  + C_{\chi\chi}\chi^2 
\\
& + A_{\psi\mathcal{K} }\dot \psi \dot{\mathcal{K}}
+ B_{\psi\mathcal{K} }\psi'\mathcal{K}' 
+ C_{\psi\mathcal{K} }\psi\mathcal{K}
+ D_{\psi\mathcal{K} }\psi'\mathcal{K}
+ E_{\psi\chi}\dot{\psi}\chi' + F_{\psi\chi}\dot{\psi}\chi
+ E_{\mathcal{K}\chi}\dot{\mathcal{K}}\chi' + F_{\mathcal{K}\chi}\dot{\mathcal{K}}\chi .
\end{split}
\label{LPMD4evenL2fin}
\end{equation}
In order to derive the dynamics of the physical degrees of freedom, one further needs to integrate out the field $\chi$. This can be done using different strategies. For instance, one can combine the equations of motion obtained from \eqref{LPMD4evenL2fin} to eliminate  the radial  derivatives acting on $\chi$ and find an algebraic equation for  $\chi$ that can be easily inverted. This is the approach  that we will adopt in this section. Alternatively, as we discuss  in detail in App.~\ref{app:quadactions}, one can instead introduce an auxiliary field at the level of the Lagrangian \eqref{LPMD4evenL2fin} which can be used to integrate out other two field components.

Let us start computing the equations of motion from \eqref{LPMD4evenL2fin}. After taking the variation with respect to the various fields and  straightforward    manipulations, one  can derive the linear superposition that is algebraic in $\chi$. Then, solving for $\chi$ and plugging the solution back into the other equations, one finds the following system:
\begin{subequations}
\label{PMevenL2}
\begin{tcolorbox}[colframe=white,arc=0pt,colback=greyish2] \vspace{-0.4cm}
\begin{align}
\psi'' + a_1 \psi' + (a_2 \omega^2 +a_3) \psi + a_4 \mathcal{K}' + (a_5 \omega^2+ a_6)\mathcal{K}  & =0 \, ,
\\
\mathcal{K}'' + b_1 \mathcal{K}' + (b_2 \omega^2 +b_3) \mathcal{K} + b_4 \psi' + (b_5 \omega^2+ b_6) \psi  & =0 \, ,
\end{align}
\end{tcolorbox}
\end{subequations}
\noindent where the prime denotes simple derivatives with respect to $r$ and where  the   coefficients $a_i$ and $b_i$ are given in App.~\ref{app:pmcoeffs}.\footnote{As opposed to \cite{Brito:2013yxa}, the coefficients $a_i$ and $b_i$ are functions of $r$ only and the frequency only appears with positive powers in the potential.}
Starting from \eqref{PMevenL2}, one can show that in the limit $\Lambda\rightarrow0$ the dynamics of the two degrees of freedom decouple and one correctly reproduces the equations of motion for a massless spin-$1$ and a massless spin-$2$ in a Schwarzschild spacetime. We will come back to this point in  App.~\ref{app:quadactions} where we show how to write the equations in a form in which  the decoupling is manifest.

 Let us focus on even modes with $L=1$. As already noticed above, when $L=1$ the perturbation $G$ automatically drops from the  parametrization \eqref{hPMeven4D}. Thus, we can still use the result \eqref{LPMD4evenL2fin} that we previously derived, except that now we still have the   gauge freedom  to eliminate another component from the partially massless field. 
One simple option is to fix $\mathcal{K}=\frac{3  }{2 \left(9 M+\Lambda  r^3\right)}\psi$ in such a way to remove the operator $\psi'^2$ from the Lagrangian.  As a result, Eq.~\eqref{LPMD4evenL2fin}  reduces to
\begin{equation}
L  =\tilde A_{\psi\psi }\dot{\psi}^2   + \tilde C_{\psi\psi } \psi^2 
+ \tilde B_{\chi\chi }\chi'^2  + \tilde C_{\chi\chi}\chi^2 
+\tilde  E_{\psi\chi}\dot{\psi}\chi'  + \tilde  F_{\psi\chi}\dot{\psi}\chi \, ,
\label{LPMD4evenL1fin}
\end{equation}
where
\begin{subequations}
\label{PMevenL1coeffs}
\begin{align}
\tilde A_{\psi\psi }  & = \frac{81 \Lambda  r^6 \left(\Lambda  r^2 (4 r+r_s)+9 r_s\right)^2}{2 \left(\Lambda  r^2+3\right) \left(\Lambda  r^3-3 r+3 r_s\right)^2 \left(2 \Lambda  r^3+9 r_s\right)^4} \, ,
\\
\tilde C_{\psi\psi }  & =  \frac{9 \Lambda  r^3 r_s^2 \left(\Lambda  r^2-9\right)^2}{\left(\Lambda  r^3-3 r+3 r_s\right) \left(2 \Lambda  r^3+9 r_s\right)^4} \, ,
\\
\tilde B_{\chi\chi }  & =  \frac{2 \Lambda  r^2}{\Lambda  r^2+3} \, ,
\\
\tilde C_{\chi\chi}  & = \frac{12 \Lambda  r \left(\Lambda  r^2+\Lambda r_s r -3\right)}{\left(\Lambda  r^2+3\right)^2 \left(\Lambda  r^3-3 r+3 r_s\right)} \, ,
\\
\tilde  E_{\psi\chi}  & =  \frac{18 \Lambda  r^4 \left(4 \Lambda  r^3+r_s \left(\Lambda  r^2+9\right)\right)}{\left(\Lambda  r^2+3\right) \left(\Lambda  r^3-3 r+3 r_s\right) \left(2 \Lambda  r^3+9 r_s\right)^2} \, ,
\\
\tilde  F_{\psi\chi}  & = \frac{36 \Lambda ^2 r^5 (2 r+r_s)}{\left(\Lambda  r^2+3\right) \left(\Lambda  r^3-3 r+3 r_s\right) \left(2 \Lambda  r^3+9 r_s\right)^2}  \, .
\end{align}
\end{subequations}
Then, integrating out $\chi$  from \eqref{LPMD4evenL1fin}, one finds
\begin{tcolorbox}[colframe=white,arc=0pt,colback=greyish2]
\begin{equation}
\frac{\D^2}{\D r_\star^2} \tilde{\psi} +  \left( \omega^2 - V_{\tilde{\psi}} \right) \tilde{\psi} =0 \, ,
\label{PMevenL1f}
\end{equation}
\end{tcolorbox}
\noindent where we defined
\begin{equation}
\psi \equiv \frac{\left(2 \Lambda  r^3+9 r_s\right)^2}{4 r \left(9-\Lambda  r^2\right)} \tilde{\psi} \, ,
\end{equation}
and where the potential is
\begin{equation}
 V_{\tilde\psi}
= 2 f \frac{\Lambda  r^2 (\Lambda  r (3 r+4 r_s)+54)-81}{r^2 \left(\Lambda  r^2-9\right)^2} \, .
\label{potevenL1PM}
\end{equation}\\
\\

\subsection{Isospectrality for partially massless spin-2 fields in 4D}
\label{sec:isoPM}

In the case of partially massless spin-$2$ fields in $4D$-SdS spacetimes, a numerical computation of the quasi-normal frequencies  has shown evidence of isospectrality \cite{Brito:2013yxa}.  In the following,  we  make  progress along this direction. First, we prove analytically isospectrality between modes with $L=1$, providing the explicit form of the symmetry transformation relating even and odd sector. This generalizes the  Chandrasekhar relation \cite{1975RSPSA.343..289C,10.2307/78902,Chandrasekhar:1985kt} to partially massless spin-$2$ fields in SdS. Second,  we discuss  the high-multipole limit, $L\gg1$, and comment on the fact that the equations of motion become degenerate, resulting in identical QNMs.

\subsubsection{Isospectrality for partially massless modes with $L=1$}

Let us  focus on the equations \eqref{massiveoddL14D} and \eqref{PMevenL1f}, describing the dynamics of partially massless modes  with $L=1$. To conform with the notation of \eqref{GDT} before, we shall identify $\Psi_+ \equiv \tilde{\psi}$, $\Psi_-\equiv Q$, $W_+\equiv \omega^2-V_{\tilde{\psi}}$ and $W_-\equiv \omega^2 - V_Q^{(L=1)} $. Then, it is not hard to check that  the potentials \eqref{VoddL1} and \eqref{potevenL1PM} are supersymmetric partners, identically satisfying the condition \eqref{GDTc2-2}. This means that the  Schr\"odinger-like equations \eqref{massiveoddL14D} and \eqref{PMevenL1f} are related each other via a Darboux transformation  in the standard form \eqref{GDT2--}. Solving \eqref{GDT3}  with $\beta \equiv 1$, or using equivalently \eqref{DTFsol}, we find
\begin{tcolorbox}[colframe=white,arc=0pt,colback=greyish2]
\begin{equation}
F(r_\star(r)) =   \frac{18 f(r)}{9 r-\Lambda  r^3} + \frac{2}{9}r_s \Lambda \, .
\label{FPMDT}
\end{equation}
\end{tcolorbox}
\noindent The function $F(r_\star(r))$ in \eqref{FPMDT} plays the role of the superpotential, which allows to rewrite the potentials for the partially massless modes with $L=1$ in the following compact form:
\begin{tcolorbox}[colframe=white,arc=0pt,colback=greyish2]
\begin{equation}
W_\pm - \omega^2 =  - F^2 \mp  \frac{\D F}{\D r_\star} + \frac{4 \Lambda ^2 r_s^2}{81} \, .
\label{FPMDTsuperp}
\end{equation}
\end{tcolorbox}
\noindent The existence of the solution \eqref{FPMDT} and the fact that it is regular at  the SdS boundaries (it asymptotes the constant $\frac{2}{9}r_s \Lambda$) are enough to show that partially massless modes with $L=1$ of different parity are isospectral.  This provides an analytic proof of the numerical findings of \cite{Brito:2013yxa} when $L=1$.
Eqs.~\eqref{FPMDT} and \eqref{FPMDTsuperp} represent therefore a novel example of superpartner  potentials for spinning particles in SdS spacetime, in addition to the massless spin-$1$ and spin-$2$ cases known so far. Note that the symmetry \eqref{FPMDT} does not have a straightforward generalization to  generic massive spin-2 fields, where, as opposed to partially massless fields, the even sector for modes with $L=1$ contains two coupled equations, which do not appear to be related to the odd equation via any  symmetry transformation.
Note, in addition, that  \eqref{FPMDTsuperp} holds formally also on SAdS spacetimes. However, the transformation \eqref{GDT2--} with $F$ given by \eqref{FPMDT}     does not induce isospectrality, because it does not preserve the AdS boundary condition \cite{Cardoso:2001bb,Berti:2009kk}.

\subsubsection{Isospectrality  in the eikonal limit}

Regarding modes with $L>1$, one can in principle generalize the dictionary of  Sec.~\ref{sec:DT} to cases in which each sector is comprised of two (coupled) equations, instead of a single one. The functions $\beta $ and $F$ will be promoted to $r$-dependent matrices, to  account for the coupling between the equations, and the condition \eqref{GDTc2} will become a set of differential equations for the various matrix components. In the case of partially massless fields with $L>1$, one can check explicitly that a simple  Darboux transformation of the type \eqref{GDT2--} where $F$ is now a $2\times 2$-matrix is not enough to relate the even and odd sector equations, at least in the form in which we have presented them here. Instead, one should  solve complicated higher-order differential equations for non-constant $\beta$, for which an analytic  solution is not known for generic values of the cosmological constant $\Lambda$. We will come back to this point in the conclusions where we will mention a possible way to overcome this issue, that we leave for future work.
There is a limit in which however the equations simplify significantly. This is the eikonal limit, which  corresponds to taking $L\rightarrow\infty$. 
In this approximation, the solution to the QNM problem  can be thought of  as a wave packet localized at the maximum of the potential, which, at the leading order in $L\rightarrow\infty$, coincides with the position of the black hole light ring.\footnote{We will discuss a similar approximation in Sec.~\ref{sec:extremal}, where we will study the extremal limit of a higher dimensional SdS black hole background. Also in  that case the dynamics is localized near the peak of the potential, resulting in analytically solvable   equations of motion for the propagating modes.}
In this sense, the  leading order of the eikonal limit is usually not very enlightening, as the dynamics  becomes independent of the spin  of the particles and of the details of the potential.\footnote{This is what allows, on the other hand, to  find an analytic expression for the QNMs in this limit.} In many cases, at this order,  isospectrality follows, therefore, trivially. 
In  the following,  we will work perturbatively in the large-$L$  limit, but we will keep subleading corrections, up to  $\mathcal{O}(L^{-1})$. At this order, the dynamics is in fact not trivial and isospectrality becomes, in principle, not obvious. In fact, we will check that the equations for partially massless modes in the even and odd sectors form two identical systems of coupled differential equations. This will automatically guarantee isospectrality up to $\mathcal{O}(L^{-1})$.

To this end, we will use the results of App.~\ref{app:quadactions}. In particular, we shall focus on the  quadratic actions \eqref{app:quadacodd} and \eqref{app:quadaceven} with $m^2=\frac{2\Lambda}{3}$.  Keeping all the terms up to the  order $\mathcal{O}(L^{-1})$, one can check that \eqref{app:quadacodd} and  \eqref{app:quadaceven} are literally mapped one into the other by
\begin{equation}
\Phi_1\rightarrow- \Psi_2\, , \qquad\qquad  \Phi_2\rightarrow \Psi_1 \, .
\label{eikrep}
\end{equation}
In particular, the equations for the parity-even modes take on the form\footnote{Recall that in the eikonal limit $L\rightarrow\infty$, the frequency $\omega$ should be considered scaling as $\omega\sim L$.}
\begin{subequations}
\label{eikonaleqseven}
\begin{align}
\frac{\D^2}{\D r_\star^2}\Phi_1 + \left[\omega^2  -f\left( \frac{L(L+1)}{r^2} - \frac{4\Lambda}{3}\right)\right] \Phi_1  & = - \frac{i \omega}{L} \sqrt{\frac{2\Lambda}{3}}  (2-L^{-1}) f \Phi_2 \, ,
\\
\frac{\D^2}{\D r_\star^2}\Phi_2 + \left[\omega^2  -f\left( \frac{L(L+1)}{r^2} - \frac{3r_s}{r^3}+ \frac{2\Lambda}{3}\right)\right] \Phi_2 & =
\frac{i \omega}{L} \sqrt{\frac{2\Lambda}{3}}  (2-L^{-1}) f \Phi_1  \, ,
\end{align}
\end{subequations}
whereas the equations for the odd-type degrees of freedom can be read off from \eqref{eikonaleqseven} via \eqref{eikrep}.
This means that, up to $\mathcal{O}(L^{-1})$, the dynamics of the even and odd degrees of freedom is described by the same system of coupled differential equations. Since the boundary conditions for the  QNMs are the same in the even and odd sectors, this  is sufficient to conclude that there is a degeneracy in the spectra at this order.
Even if isospectrality is not a trivial result at the order in the eikonal limit that we are considering here, we note, however, that it is perhaps not as remarkable as in the case discussed above.
Indeed, as opposed to the $L=1$ result  \eqref{FPMDTsuperp}, which crucially depends on the precise form of the potentials and does not rely on any approximations, in the eikonal limit there are still wide classes of potentials that yield degenerate quasi-normal frequencies \cite{Cardoso:2019mqo}. In this sense, the isospectrality in the large-$L$ limit does not fully make manifest what is special about partially massless fields, even though it still provides a non-trivial check of our results against \cite{Brito:2013yxa}.\footnote{Note  that, at the order $\mathcal{O}(L^{-1})$, there is nevertheless a neat distinction between the dynamics of massive and partially massless fields. In particular, our argument above does not straightforwardly apply  if $m^2\neq\frac{2\Lambda}{3}$. Indeed, the equations \eqref{massiveeqsD4} describing the dynamics of the parity-even components are still non-trivially coupled and there is no manifest duality symmetry---like the one in \eqref{eikrep}---with the odd sector.}

 Note  that this result is  exclusive of four spacetime dimensions.  In the next section, we will see indeed that it does not remain true in higher-dimensional SdS spacetimes, where partially massless modes of opposite parity cease to be isospectral.

\vspace{1cm}

\section{Massive and partially massless spin-2 fields on S(A)dS in arbitrary dimensions}
\label{sec:massiveD}

In this section we extend the results of the previous sections for massive spin-$2$ fields on S(A)dS spacetimes to arbitrary dimensions. A complete and pedagogical introduction to perturbation theory around a  S(A)dS background in general dimensions can be found in \cite{Hui:2020xxx}.    In the following, we will consider in particular partially massless perturbations and check explicitly that isospectrality is broken for $D\neq4$.  Even though this result is not unexpected, to the best of our knowledge it has not been presented previously in the literature.   Furthermore,  we will discuss the stability of massive spin-2 perturbations in S(A)dS backgrounds and check for the presence of a Gregory-Laflamme instability \cite{Gregory:1993vy,Babichev:2013una} (see also \cite{Brito:2013wya}) of the monopole-type. 

Our starting point is the quadratic Fierz-Pauli action for a spin-$2$ field $h_{\mu\nu}$ of generic mass $m$ now generalized to a $D$-dimensional S(A)dS spacetime,
\begin{multline}
S=\int \D^Dx\sqrt{-g}\left[ -{1\over 2}\nabla_\lambda h_{\mu\nu} \nabla^\lambda h^{\mu\nu}+\nabla_\lambda h_{\mu\nu} \nabla^\nu h^{\mu\lambda}-\nabla_\mu h\nabla_\nu h^{\mu\nu}+\frac{1}{2} \nabla_\mu h\nabla^\mu h 
\right.
\\
\left.
+\frac{R}{D}\left( h^{\mu\nu}h_{\mu\nu}-\frac{1}{2} h^2\right)
-\frac{1}{2}m^2\left( h^{\mu\nu}h_{\mu\nu}- h^2\right) 
\right] \, .
\label{Sspin-2massive}
\end{multline}
The field $h_{\mu\nu}$ is again assumed to propagate on a fixed S(A)dS background, whose metric $g_{\mu\nu}$ is given by
\begin{equation}
\D s^2 = g_{\mu\nu}\D x^\mu\D x^\nu = -f(r) \D t^2 + \frac{1}{f(r)}\D r^2+r^2 \D\Omega_{S^{D-2}}^2,
\label{eq:sphericalmetric}
\end{equation}
where
\begin{equation}
f(r)  = 1-\left(\frac{r_s}{r}\right)^{D-3}-\frac{2\Lambda r^2}{(D-1)(D-2)M_{\rm Pl}^2} ,
\label{eq:sadsf}
\end{equation}
which satisfies the background Einstein equations,
\begin{equation}
R_{\mu\nu} = \frac{R}{D} g_{\mu\nu} \, ,
\qquad
\Lambda = \frac{D-2}{2D}R \, .
\end{equation}
The cosmological constant $\Lambda$ is defined such that  the Einstein-Hilbert action takes the form $\frac{1}{2}M_{\rm Pl}^{D-2}(R-2\Lambda)$ with $\delta g_{\mu\nu} = 2M_{\rm Pl}^{-\frac{D-2}{2}}h_{\mu\nu}$ \cite{Hinterbichler:2011tt}.

Given the spherical symmetry of the background, we now decompose  and classify the metric perturbations according to  the $SO(D-1)$ rotational symmetry group.  Thus, we  write \cite{Kodama:2003jz,Ishibashi:2003ap,Hui:2020xxx}  
%
\begin{subequations}
\label{spin2decomposition}
\begin{align}
h_{tt} &= \sum_{L,M}  Y_L^M(\theta) f(r)H_0(t,r), \\
h_{tr} &= \sum_{L,M}  Y_L^M(\theta) H_1(t,r),\\
h_{rr} &= \sum_{L,M}  Y_L^M(\theta) f^{-1}H_2(t,r),\\
h_{ti} &= \sum_{L,M}\left(\nabla_i Y_L^M(\theta) {\cal H}_0(t,r)+   Y_i^{(T)}{}_L^M(\theta) h_0(t,r)\right),\\
h_{ri} &=\sum_{L,M}\left(\nabla_i Y_L^M(\theta) {\cal H}_1(t,r)+Y_i^{(T)}{}_L^M(\theta) h_1(t,r)\right), \\
h_{ij} &= \sum_{L,M}r^2\Big(\gamma_{ij} Y_L^M(\theta) {\cal K}(t,r)+\nabla_{(i}\nabla_{j)_T}Y_L^M(\theta) G(t,r)
\notag \\
&\qquad\qquad\qquad\qquad+\nabla_{(i}Y_{j)}^{(T)}{}_L^M(\theta) h_2(t,r)+ Y^{(TT)}_{ij}{}_L^M(\theta) h_T(t,r)
\Big) \, ,
\end{align}
\end{subequations}
where $\gamma_{ij}$ is the induced metric on the $(D-2)$-sphere $S^{D-2}$ and the Latin indices $i, j, k\ldots$ denotes coordinates on $S^{D-2}$.   $Y_L^M(\theta)$, $Y_i^{(T)}{}_L^M(\theta)$ and $Y^{(TT)}_{ij}{}_L^M(\theta)$ are scalar, (transverse) vector and (transverse-traceless) tensor (hyper-)spherical harmonics.\footnote{Further details can be found in \cite{Hui:2020xxx}. Note that our notation and conventions here exactly mirror the ones in \cite{Hui:2020xxx}, except however for the vector harmonics $Y_i^{(T)}{}_L^M(\theta)$, which we take normalized to $L(L+D-3)$, instead of unity as in \cite{Hui:2020xxx}.}
$M$ is a multi-index cataloguing the various magnetic quantum numbers \cite{Hui:2020xxx}.  
In \eqref{spin2decomposition}, $( \, \cdots)_T$ denotes the trace-free symmetrized part of the enclosed indices, e.g.,
\begin{equation}
\nabla_{(i}\nabla_{j)_T}Y_L^M(\theta) = \frac{1}{2}\left( \nabla_{i}\nabla_j + \nabla_{j}\nabla_i - \gamma_{ij} \nabla_k\nabla^k \right) Y_L^M(\theta) \, .
\end{equation}
Note that the tensor spherical harmonic $Y_{ij}^{(TT)}$  is not present in $D=4$, but is special of higher dimensions.

Next, we plug the decomposition \eqref{spin2decomposition} into the action \eqref{Sspin-2massive}.
As in $D=4$, the spherical symmetry of the background allows to treat modes with different transformation laws under the action of the $SO(D-1)$  group separately, since their  equations of motion are guaranteed to  decouple at the linear level.

\subsection{Tensor sector (tensor harmonics)}
\label{sec:tensor-D}

Tensor harmonics only exist in $D>4$. Following the decomposition \eqref{spin2decomposition}, the tensor sector is described by the following  spin-$2$ perturbations:
\begin{equation}
h_{tt}= h_{rr} = h_{ti} = h_{ri } =0 \, ,
\qquad
h_{ij} = r^2  h_T(t,r)  Y^{(TT)}_{ij}{}_L^M(\theta)  \, .
\label{htensor}
\end{equation}
The Lagrangian for the tensor perturbation $h_T$ takes on the form
\begin{equation}
L = \frac{ r^{D -2}}{2 f} \left[ \dot{h}_T^2 - f^2 h_T'^2 
 +   \frac{f}{r^2} h_T^2   \left( 2 r f'+2 (D -3) (f-1)-L(L+D-3) -m^2 r^2+\frac{4 \Lambda  r^2}{D -2}\right) \right] \, .
\end{equation}
Taking the variation with respect to $h_T$ and defining 
\begin{equation}
h_3 \equiv r^{1-\frac{D}{2}} h_T \, ,
\end{equation}
we find the following Schr\"odinger-like equation:
\begin{tcolorbox}[colframe=white,arc=0pt,colback=greyish2]
\begin{equation}
\frac{\D^2}{\D r_\star^2} h_3 + \left( \omega^2 - V_T \right) h_3 =0 \, ,
\label{eqtensorD}
\end{equation}
\end{tcolorbox}
\noindent where
\begin{equation}
V_T = f \left[\frac{\left(D^2-14 D +32\right) f}{4 r^2} + \frac{(D -6)  f'}{2 r} + \frac{ \left(2 (D -3)+L^2+(D -3) L\right)}{r^2}+ m^2 - \frac{4 \Lambda }{D -2} \right] \, .
\end{equation}

\subsection{Odd sector (vector harmonics)}
\label{sec:odd-D}

Let us consider now  the odd sector. After plugging the field decomposition \eqref{spin2decomposition} into the action \eqref{Sspin-2massive}, one finds the following Lagrangian for the odd components:
\begin{multline}
L = r^{D-4} \bigg[
\dot{h}_1^2 +\frac{4 }{r} h_0 \dot{h}_1 
-\frac{(L-1) (D +L-2) }{f} h_0 \dot{h}_2
+f (L-1)  (D +L-2) h_1 h_2'
-2 h_0'  \dot{h}_1 
+h_0'^2
\\
+\frac{1}{4} (L-1) (D +L-2) \frac{r^2 \dot{h}_2^2}{f}
-\frac{1}{4} (L-1) (D +L-2) r^2 f h_2'^2
\\
\frac{ -2 r f'+(L+1) (D +L-4)+r^2 \left(m^2-\frac{4 \Lambda }{D -2}\right)}{r^2 f} h_0^2
\\
+\frac{f  \left(2 r f'+2 (D -3) f-(L+1) (D +L-4)+r^2 \left(\frac{4 \Lambda }{D -2}-m^2\right)\right)}{r^2} h_1^2
\\
+\frac{1}{4} (L-1) (D +L-2) \left(2 r f'  +2 (D -3) f  -\frac{r^2 ((D -2) m^2-4 \Lambda ) }{D -2}-2 (D -3)\right)h_2^2
\bigg] \, .
\label{oddactionD}
\end{multline}
For generic values of $L\geq2$, after taking the variation with respect $h_0$, $h_1$ and $h_2$, and simple algebraic manipulations, one can  integrate out $h_0$ and find the following system of coupled differential equations:
\begin{subequations}
\begin{tcolorbox}[colframe=white,arc=0pt,colback=greyish2] \vspace{-0.4cm}
\begin{align}
\frac{\D^2}{\D r_\star^2} Q +  \left( \omega^2 - V_Q \right) Q  & = S_Z Z \, ,
\\
\frac{\D^2}{\D r_\star^2} Z +  \left( \omega^2 - V_Z \right) Z  & = S_Q Q \, ,
\end{align}
\end{tcolorbox}
\label{PModdL2}
\end{subequations}
\noindent where we have defined the fields
\begin{subequations}
\begin{align}
Q &  \equiv - f^{-1/2}r^{-D -2} \left(2 \Lambda  r_s^3 r^{D +2}-\left(D ^2-3 D +2\right) r_s^3 r^{D }+\left(D ^2-3 D +2\right) r^3 r_s^{D }\right)^{3/2} h_1 \, ,
\label{Scspin2oddD-1}
\\
Z &  \equiv -3 f^{-1/2} r_s^3 r^{-1 }\left(2 \Lambda  r_s^3 r^{D +2}-\left(D ^2-3 D +2\right) r_s^3 r^{D }+\left(D ^2-3 D +2\right) r^3 r_s^{D }\right)^{1/2} h_2 \, ,
\label{Scspin2oddD-2}
\end{align}
\label{Scspin2oddD}
\end{subequations}
and where
\begin{subequations}
\label{Scspin2oddDpot}
\begin{align}
V_Q &=  \frac{r^{-2 (D+1)}}{4 (D-2) (D-1)^2 r_s^6} 
 \left[\left(D^2-3 D+2\right) r_s^3 r^{D}-\left(D^2-3 D+2\right) r^3 r_s^{D}-2 \Lambda  r_s^3 r^{D+2}\right]
\notag\\
&\quad
	\cdot \bigg[(D-1) r_s^3 \left(D^2+D (4 L-2)+4 \left(L^2-3 L+2\right)\right) r^{D}-D \left(3 D^2-7 D+4\right) r^3 r_s^{D}
\notag\\
&\quad	\qquad
	+2 r_s^3 r^{D+2} (2 (D-1) m^2-D \Lambda )
\bigg] \, ,
\label{VQoddD}
\\
V_Z &=  \frac{r^{-2 (D+1)}}{4 (D-2) (D-1)^2 r_s^6}
	 \left[\left(D^2-3 D+2\right) r_s^3 r^{D}-\left(D^2-3 D+2\right) r^3 r_s^{D}-2 \Lambda  r_s^3 r^{D+2}\right]
\notag\\
&\quad \cdot \bigg[(D-1) r_s^3 \left(D^2+2 D (2 L-5)+4 \left(L^2-3 L+4\right)\right) r^{D}
+(D-2)^2 (D-1) r^3 r_s^{D}
\notag\\
&\quad	\qquad
	+2 r_s^3 r^{D+2} (2 (D-1) m^2-D \Lambda )
\bigg] \, ,
\\
S_Z &=  f^2\frac{(D-2)^2 (D-1)^2 (L-1) (D+L-2)  \left((D-1) r^3 r_s^{D}-2 r_s^3 r^{D}\right)}{6 r_s^3 r^{D+2} \left(-(D-3) D+2 \Lambda  r^2-2\right)+6 (D-2) (D-1) r^5 r_s^{D}} \, ,
\\
S_Q &= - f^2 \frac{12 r_s^3}{r^2 r_s^3 \left(-(D-3) D+2 \Lambda  r^2-2\right)+(D-2) (D-1) r^{5-D} r_s^{D}}
 \, .
\end{align}
\end{subequations}
Notice that Eqs.~\eqref{Scspin2oddDpot} correctly reproduce Eqs.~\eqref{VSodd4D} in $D=4$.

The equation for the odd modes with $L=1$ can be easily obtained by setting  $L=1$ into the action \eqref{oddactionD}. As expected, the perturbation $h_2$ drops out and one is left with a Lagrangian for $h_0$ and $h_1$ only. Integrating out the former yields
\begin{tcolorbox}[colframe=white,arc=0pt,colback=greyish2] 
\begin{equation}
\frac{\D^2}{\D r_\star^2} Q +  \left( \omega^2 - V_Q^{(L=1)} \right) Q =0 \, ,
\label{massiveoddL1D}
\end{equation}
\end{tcolorbox}
\noindent where $Q$ is defined as in \eqref{Scspin2oddD-1} while the potential $V_Q^{(L=1)}$ is given by Eq.~\eqref{VQoddD} with $L=1$.

\subsection{Even sector (scalar harmonics)}

The derivation of the equations for the parity-even perturbations  in arbitrary dimensions  is a straightforward generalization of the procedure outlined in \eqref{subsec:evenmassive4D}. The main difference is that the coefficients in the action will now depend on the number of dimensions $D$.   Let us briefly streamline the derivation  starting with the modes with $L\geq2$ and highlight the main expressions.  The constraint equation enforced by $H_0$ now reads
\begin{multline}
\mathcal{H}_1 (D -2) \left(2  L r (D +L-3) f'+4 L(D -3)   (D +L-3)f\right)
\\
+\mathcal{H}'_1 \left(4 (D -2) L^2 r f-12 (D -2) L r f+4 (D -2) D  L r f\right)
-2H_2' (D -2)^2 r^2 f 
\\
+ H_2 r(D-2) \left(  r^2 f'' -(D -2) r f'-2 (D -3) (D -2) f-2   L^2 -2  L(D-3) -2   m^2 r^2  + \frac{4 \Lambda  r^2}{D-2} \right)
\\
+ \mathcal{K} (D -2)^2 \left(2  (D-3) r f  +  2  r \left( r f'- \frac{D-3}{D-2} L(L+D-3)-r^2  m^2+ \frac{2r^2 \Lambda}{D-2} \right)\right)
\\
+ \mathcal{K}' (D -2) \left((D -2) r^3 f'+2 D^2 r^2 f+4  r^2 f-6  D  r^2 f\right)
\\
+ 2\mathcal{K}''r^3 f (D -2)^2
\\
-2G  (D -3) (L-1) L r \left((D-2)(D-3) +L(L+2 D -5) \right)  =0 \, .
\label{constrH0D}
\end{multline}
Again, it is convenient to introduce the field $\psi$, defined by
\begin{equation}
H_2(t,r) \equiv \frac{2 L (D +L-3) }{(D -2) r} \mathcal{H}_1(t,r)  -\frac{r^{3-D }}{(D -2) f}  \psi(t,r) +r \mathcal{K}'(t,r) \, ,
\label{shiftH2D}
\end{equation}
in such a way to be able to solve \eqref{constrH0D} algebraically in $\mathcal{H}_1$:
\begin{multline}
\mathcal{H}_1
= 
\frac{r^{2}(D-2) \left( 2 L(L+D-3) +2r^2  m^2 -\frac{4r^2 \Lambda}{D-2} - r^2 f'' -2 (D -2) f\right)   \mathcal{K}'  }{2 L  (L+D-3) \left( r^2 f''+2 (D -2) f-2  L(L+D-3) -2r^2 m^2+\frac{4r^2 \Lambda}{D-2} \right)}
\\
+\frac{ r(D -2)  \left( (D -3) L(L+D-3)+r^2 ((D -2) m^2 -2 \Lambda )- (D -3) (D -2) f -(D -2) r f' \right)  \mathcal{K} }{ L  (L+D-3) \left( r^2 f''+2 (D -2) f-2  L(L+D-3) -2r^2 m^2+\frac{4r^2 \Lambda}{D-2} \right)}
\\
-\frac{(D -2) r^{5-D } \psi'}{L (L+D -3) \left( r^2 f''+2 (D -2) f-2  L(L+D-3) -2r^2 m^2+\frac{4r^2 \Lambda}{D-2} \right)}
\\
+\frac{r^{4-D }   \left( r^2 f''+(D -2) r f'-2 \left( L(L+D-3)+r^2 m^2 -\frac{2r^2 \Lambda}{D-2} \right)\right)   \psi  }{2 L f (D +L-3) \left( r^2 f''+2 (D -2) f-2  L(L+D-3) -2r^2 m^2+\frac{4r^2 \Lambda}{D-2} \right)}
\\
+ \frac{(D -3)  (L-1) r \left((D-2)(D-3)+L(L+2 D -5) \right)  G }{(D +L-3) \left( r^2 f''+2 (D -2) f-2  L(L+D-3) -2r^2 m^2+\frac{4r^2 \Lambda}{D-2} \right)}  \, .
\end{multline} 
After integrating out  $H_1$ and shifting the field $\mathcal{H}_0$ as follows,
\begin{multline}
\mathcal{H}_0 =  \chi
-\frac{(D -2) r^2}{2 L (D +L-3)}  \dot{\mathcal{K}}
\\
+\frac{(D -2) r^5 r_s^3 \dot{\psi}}{L (D +L-3) \left(2 (L-1) r_s^3 (L+D-2) r^{D }+2 m^2 r_s^3 r^{D +2}+\left(D ^2-3 D +2\right) r^3 r_s^{D }\right)}
\end{multline}
in such a way to get rid of the higher derivative terms $\dot{\psi}'^2$ and  $\dot{\mathcal{K}}'^2$, the final Lagrangian take the form\footnote{Again, as in \eqref{action4DMassive}, we are setting $M=0$. This will not affect the resulting equations of motion, thanks to the spherical symmetry of the background.}
\begin{equation}
\begin{split}
L & =  A_{\psi\psi }\dot\psi^2 + B_{\psi\psi }\psi'^2  + C_{\psi\psi }\psi^2 
+ A_{\mathcal{K}\mathcal{K} }\dot{\mathcal{K}}^2 + B_{\mathcal{K}\mathcal{K} }\mathcal{K}'^2  + C_{\mathcal{K}\mathcal{K} }\mathcal{K}^2 
\\
&\qquad
+ A_{GG }\dot G^2 + B_{GG }G'^2  + C_{GG }G^2 
+ B_{\chi\chi }\chi'^2  + C_{\chi\chi}\chi^2 
\\
&\qquad + A_{\psi\mathcal{K} }\dot \psi \dot{\mathcal{K}}
+ A_{\psi G} \dot \psi \dot{G}
+ A_{G\mathcal{K} }\dot G \dot{\mathcal{K}}
+ B_{\psi\mathcal{K} }\psi'\mathcal{K}' 
+ B_{\psi G }\psi' G'
+ B_{G\mathcal{K} }G'\mathcal{K}' 
\\
&\qquad
+ C_{\psi\mathcal{K} }\psi\mathcal{K}
+ C_{\psi G }\psi G
+ C_{G\mathcal{K} }G\mathcal{K}
+ D_{\psi\mathcal{K} }\psi'\mathcal{K}
+ D_{\psi G }\psi' G
+ D_{\mathcal{K}G }\mathcal{K}' G
\\
&\qquad
+ E_{\psi\chi}\dot{\psi}\chi' + F_{\psi\chi}\dot{\psi}\chi
+ E_{\mathcal{K}\chi}\dot{\mathcal{K}}\chi' + F_{\mathcal{K}\chi}\dot{\mathcal{K}}\chi
\\
&\qquad
+ E_{G\chi}\dot{G}\chi' + F_{G\chi}\dot{G}\chi 
 \, .
\end{split}
\label{LPMevenL2fin}
\end{equation}
Since the expressions for the coefficients in \eqref{LPMevenL2fin} are quite lengthy, we will provide them separately in an ancillary file.

Given the action \eqref{LPMevenL2fin} one can further integrate out the field $\chi(t,r)$ and derive a system of $3$ coupled differential equations for the $3$  propagating degrees of freedom in the even sector. The expressions for the equations are quite complicated and we will not write them explicitly. Instead, we will discuss below some particular cases. In particular, we will study the extremal limit for massive spin-$2$ perturbations in arbitrary dimensions and discuss the partially massless case, showing explicitly the breaking of isospectrality when $D\neq 4$.

However, before getting there, we still have to check the  modes with $L=1$ and $L=0$.
In the former case, one can consistently start from the Lagrangian \eqref{LPMevenL2fin} and set $L=1$. The field component $G$ will drop from the action and one is left with a system of two coupled differential equations for $\psi$ and $\mathcal{K}$---this case is formally analogous to the case of  partially massless fields with $L\geq2$, therefore we refer to Sec.~\ref{subsec:evenmassive4D} for further details.

The case $L=0$, on the contrary, can not be inferred straightforwardly from \eqref{LPMevenL2fin}. Instead, we need to revisit our derivation starting again from  \eqref{Sspin-2massive}.   When $L=0$, the field components $\mathcal{H}_0$, $\mathcal{H}_1$ and $G$ drop from the parametrization \eqref{spin2decomposition}. In the action \eqref{Sspin-2massive}, the field $H_0$ is still a Lagrange multiplier. However, in the present case, instead of shifting $H_2$ as in \eqref{shiftH2D}, it is convenient to  introduce the following field redefinition,
\begin{equation}
H_2 (t,r) 
\equiv   -\frac{r^{3-D }}{(D -2) f}  \psi(t,r) +r \mathcal{K}'(t,r)
+ \frac{ 2 (D -2) f +  f''  - 2  m^2 + \frac{4 \Lambda}{D-2} }{2 (D -2) f}r^2 \mathcal{K}(t,r) \, ,
\label{H2L0D}
\end{equation}
in such a way to remove all the derivatives acting on $\mathcal{K}$ from the $H_0$'s equation of motion. Then, solving algebraically for $\mathcal{K}$ yields
\begin{multline}
\mathcal{K} =
-\frac{2 (D -1) r r_s^3  }{2 (D -1) m^2 r_s^3 r^{D }-4 \Lambda  r_s^3 r^{D }+\left(D ^3-6 D ^2+11 D -6\right) r r_s^{D }} \psi
\\
+ \frac{2 r^{-D } \left(2 \Lambda  r_s^3 r^{D +2}-\left(D ^2-3 D +2\right) r_s^3 r^{D }+\left(D ^2-3 D +2\right) r^3r_s^{D }\right)}{m^2 \left(2 (D -1) m^2 r_s^3 r^{D }-4 \Lambda  r_s^3 r^{D }+\left(D ^3-6 D ^2+11 D -6\right) r r_s^{D }\right)} \psi'  \, .
\label{KL0D}
\end{multline}
After integrating out $H_1$ and plugging its solution back into the action, along with the expressions \eqref{H2L0D} and \eqref{KL0D}, one finds a Lagrangian  for $\psi $ only. Removing all the higher derivative terms by simple integrations by parts, the Lagrangian takes on the standard form with  kinetic terms and quadratic potential. The corresponding Schr\"odinger-like  equation of motion is
\begin{tcolorbox}[colframe=white,arc=0pt,colback=greyish2] 
\begin{equation}
\frac{\D^2}{\D r_\star^2}\phi + \left( \omega^2 - V_\phi \right)\phi = 0 \, ,
\label{scalarD}
\end{equation}
\end{tcolorbox}
\noindent where 
\begin{equation}
\phi \equiv \frac{\left(D ^2-3 D +2\right)^{1/2} r_s^{3/2} r^{D/2 }}{ 2 (D -1) m^2 r_s^3 r^{D }-4 \Lambda  r_s^3 r^{D }+\left(D ^3-6 D ^2+11 D -6\right) r r_s^{D }} \psi 
\end{equation}
and where the potential is
\begin{multline}
V_\phi =
-  \big[ r^{-2 (D +1)} \left((D -2) (D -1) r^3 r_s^{D }-r_s^3 r^{D } \left((D -3) D -2 \Lambda  r^2+2\right)\right)
\\
((D -3) (D -2) (D -1)^2 r^{D +2} r_s^{2 D +3} ((D -1) \left((D -4) (D -3) (D -2)^2+4 (D  (2 D -5)+6) m^2 r^2\right)
\\
-2 (D  (D  (D +7)-26)+24) \Lambda  r^2)
 -12 (D -2) (D -1) r^{2 D +1} r_s^{D +6} ((D -1) m^2 -2 \Lambda )
 \\
  \left((D -1) \left((D -3) (D -2) D +(D +2) m^2 r^2\right)
 -2 ((D -2) D +2) \Lambda  r^2\right)
 +4 r_s^9 r^{3 D } (2 \Lambda -D  m^2+m^2)^2
 \\  
  \left((D -1) \left(D  (D +2)+4 m^2 r^2\right)-2 D  \Lambda  r^2\right)+(D -3)^2 (D -2)^4 (D -1)^3 r^5 r_s^{3 D }) \big]
\\
\cdot \left[ 4 (D -2) (D -1)^2 r_s^6 \left(2 r_s^3 r^{D } ((D -1) m^2-2 \Lambda )+(D -3) (D -2) (D -1) r r_s^{D }\right)^2 \right]^{-1} .
\end{multline}
We will come back to this in the next section when we discuss the presence of a  Gregory-Laflamme instability in the spectrum of \eqref{scalarD}.

\subsection{The extremal limit: breaking of isospectrality and the Gregory-Laflamme instability}
\label{sec:extremal}

The explicit expressions for the potentials associated with massive spin-$2$ fields in arbitrary dimensions are quite complicated, especially for the even sector.  However, we can consider an extremal limit in which we increase the value of the cosmological constant so that the two horizons of the SdS spacetime coincide.  In this limit, the expressions  take a simple form and one can expand the field's potentials in powers of the relative distance \cite{Cardoso:2003sw}.  In particular, it was shown in \cite{Brito:2013yxa} that in this regime and in $D=4$ spacetime dimensions, partially massless modes are isospectral.  Here, we extend the results of \cite{Brito:2013yxa} to generic spacetime dimensions.  We show that, in fact, all massive spin-2 modes are isospectral in the extremal limit in $D=4$.  We furthermore show explicitly that this isospectrality is broken for spacetime dimensions not equal to 4, thus implying that partially massless perturbations are not isospectral beyond $D=4$, like massless spin-2 perturbations.  Furthermore, we use the extremal limit to extend the work of \cite{Babichev:2013una,Brito:2013wya} to arbitrary spacetime dimensions, showing that the helicity-0 mode of the massive spin-2 perturbation is unstable on static Schwarschild or Schwarzschild de Sitter spacetimes.  The technical details about the extremal limit in arbitrary dimensions are discussed in App.~\ref{app:extremallimit}. In the following, we will report the expressions for the potentials at the leading order in this limit and comment about the QNM spectra.

Starting from \eqref{eqtensorD}, \eqref{PModdL2} and  the system of equations obtained from the action \eqref{LPMevenL2fin} after $\chi$ is integrated out,
in the extremal limit one can easily find a way to decouple all the equations (in both odd and even sector). Eventually, they will  take a Schr\"odinger-like form with P\"oshl-Teller potential (see App.~\ref{app:extremallimit}),
\begin{tcolorbox}[colframe=white,arc=0pt,colback=greyish2] 
\begin{equation}
\frac{\D^2}{\D r_\star^2}\Psi(r) + \left(\omega^2 - \frac{\kappa (r_c-r_b)U(\bar r)}{2\cosh^2(\kappa r_\star)} \right) \Psi(r) =0 \, , 
\label{EQexDDim}
\end{equation}
\end{tcolorbox}
\noindent where $\Psi$ generically denotes the various field components and where  
\begin{equation}
U(\bar r) = \begin{cases}
m^2 + \dfrac{ L  (L+D-3)-D+2}{\bar r^2} - \dfrac{D-4}{\bar r^2}	    & \text{ (even)} \, ,
\\
m^2 + \dfrac{ L  (L+D-3)-D+2}{\bar r^2} 	& \text{ (odd)} \, ,
\\
m^2 + \dfrac{L ( L+D-3)}{\bar r^2}	& \text{ (tensor)} \, .
\end{cases}
\label{UextremalmassiveL}
\end{equation}
$\bar{r}$ represents the distance, defined in Eq.~\eqref{extremallimit}, where the two horizons coincide, while  $\kappa$ is defined in \eqref{kappadef}.
The equation \eqref{EQexDDim} is particularly convenient because it admits an analytic expression for the QNMs \cite{Ferrari:1984zz,PhysRevLett.52.1361}  (see also App.~\ref{app:extremallimit} for notation and further details): 
\begin{equation}
\frac{\omega}{\kappa} = -i \left(n + \frac{1}{2}\right) + \sqrt{\frac{\bar r^2 U(\bar r) }{D-3}-\frac{1}{4} } \, ,
\qquad
n = 0, 1\ldots
\label{QNMextremalPML1}
\end{equation}
where $n$ takes non-negative integer values.
Now, it is worth recalling that isospectrality is generically broken for massive fields in $4$ and higher dimensional SdS spacetimes \cite{Rosa:2011my,Brito:2013wya}.
However, Eq.~\eqref{UextremalmassiveL} shows that isospectrality is recovered at least in the extremal limit if $D=4$. In other words, in the extremal limit,  all the different spectra become degenerate, resulting in identical sets of quasi-normal frequencies.
Notice that this is true only in $D=4$. In higher dimensions, as it is clear from \eqref{UextremalmassiveL}, this property ceases to hold and one ends up  with  different QNMs.


The above result can be immediately specialized to the case of partially massless fields: 
the dynamics is still dictated by the single Schr\"odinger-like equation  \eqref{EQexDDim}, where $U(\bar r)$ is given by \eqref{UextremalmassiveL} with  $m^2=\frac{2\Lambda}{D-1}$.
Thus isospectrality is also broken for partially massless modes away from $D=4$.  The breaking of isospectrality away from $D=4$ for partially massless spin-2 perturbations was indeed expected, at least in the small-$\Lambda$ limit, where the vectorial and tensorial components of the partially massless perturbations decouple and reproduce the equations for massless spin-$1$ and spin-$2$ fields in arbitrary dimensions (see App.~\ref{app:quadactions} for the derivation), whose even and odd sectors are known to be non-isospectral~\cite{Konoplya:2003dd}. In this section, we have shown explicitly that this remains generically true, also away from the limit of asymptotically flat spacetimes.


\vspace{1cm}


\noindent {\bf Gregory-Laflamme Instability:} In \cite{Gregory:1993vy}, Gregory and Laflamme  showed  that black strings and black branes, the low energy solutions of string theory, are   unstable against small fluctuations. Such an instability manifests as exponentially growing solutions to the linearized equations of motion for the classical perturbations around the background configuration. A similar issue affects the longitudinal component of a massive spin-$2$ field in $D=4$ \cite{Babichev:2013una,Brito:2013wya}, indicating that massive spin-2 particles in $4$-dimensional static, Schwarzschild or SdS spacetimes are unstable.  This is perhaps unsurprising, given the arguments that physical black hole solutions in the full theory of massive gravity are likely time-dependent (see, e.g., \cite{Mirbabayi:2013sva,Rosen:2017dvn}).

In this section, we confirm this expectation, generalizing the results of \cite{Babichev:2013una,Brito:2013wya} to Schwarzschild and SdS spacetimes in arbitrary $D$-dimensions.  For simplicity, we focus again on the extremal limit, which admits an analytic expression for the QNM spectrum and allows  to easily visualize the origin of the instability.
Our starting point is Eq.~\eqref{scalarD}. In the extremal limit, it simplifies significantly, taking on the  form
\begin{tcolorbox}[colframe=white,arc=0pt,colback=greyish2] 
\begin{equation}
\frac{\D^2}{\D r_\star^2}\phi(r) + \left(\omega^2 - \frac{\kappa (r_c-r_b)U(\bar r)}{2\cosh^2(\kappa r_\star)} \right) \phi(r) =0 \, , 
\end{equation}
\end{tcolorbox}
\noindent with a P\"oshl-Teller potential, where
\begin{equation}
U(\bar r) = m^2 - \frac{2 (D -3)}{\bar r^2} \, .
\end{equation}
The  quasi-normal frequencies can be obtained by solving for fixed $n$  Eq.~\eqref{QNMextremalPML1} (see  App.~\ref{app:extremallimit} for details), which now reads
\begin{equation}
\frac{\omega}{\kappa} = -i \left(n + \frac{1}{2}\right) + \sqrt{  \frac{\bar r^2 m^2 }{D-3} -\frac{9}{4} } \, ,
\qquad
n = 0, 1\ldots
\label{QNMextremalscalar}
\end{equation}
Notice that the argument inside the square root is not always positive definite. It may happen that, for certain values of the mass, the second term in \eqref{QNMextremalscalar} becomes imaginary. In particular,  if this contribution is large enough to compensate the first term in \eqref{QNMextremalscalar}, the imaginary part of the frequency flips sign, corresponding therefore to an instability in the spectrum. In the case of $(L=0)$-modes for massive spin-$2$ fields in arbitrary dimensions, this occurs when
\begin{equation}
\frac{(D-2)(D-3)}{D-1} < m^2 \bar{r}^2 < 2(D-3) \, ,
\label{GLinstability}
\end{equation}
where $\frac{(D-2)(D-3)}{D-1}$ is the Higuchi bound for $\Lambda=\bar{\Lambda}$ (see Eq.~\eqref{extremallimit}). In the range of values given by \eqref{GLinstability}, it is guaranteed that there is  at least one unstable mode in the QNM spectrum. Notice that, in $D=4$, the right side of the inequality \eqref{GLinstability} becomes $\frac{m r_s}{2}<\frac{\sqrt{2}}{3} \approx 0.47$, in agreement with the numerical results of \cite{Brito:2013wya}. In addition, for fixed values of the mass $m^2$ in units of $r_s^2$, the instability rate grows as $D$ increases, in line with the general behaviour of the Gregory-Laflamme instability in string theory in the small-$\Lambda$ limit \cite{Gregory:1993vy}.

\section{Conclusions}
Isospectrality and its peculiarity to 4 spacetime dimensions remains a curious phenomenon.  Above we presented analytic arguments for the isospectrality of partially massless spin-2 particles in a Schwarzschild  de Sitter background in 4$D$ and the breaking of isospectrality away from $D=4$ dimensions. 
In particular, we have provided the explicit form of the symmetry that underlies  isospectrality between  modes with multipole number $L=1$ in $D=4$. This generalizes the Chandrasekhar relation for black hole perturbations in general relativity and represents a novel example of duality between modes of different parity for spinning particles on SdS backgrounds.

An analytic proof of isospectrality for partially massless fields  of  generic $L$ is still, however,  elusive.
In this work, we have checked that when $L>1$ the potentials are not related via a standard Chandrasekhar relation, but  require instead the introduction of a generalized Darboux transformation \cite{Glampedakis:2017rar}. This amounts to solving higher-order differential equations, for which an analytic solution is not easy to find for generic values of the cosmological constant. This does not mean though that one can not find in principle a field reparametrization that allows to recast the potentials in a form that makes them manifestly supersymmetric partners. It is possible that other approaches such as the Newman-Penrose formalism might shed some light on this \cite{Newman:1961qr,Bardeen:1973xb,Teukolsky:1973ha}. 
This would have several advantages. First, it may allow one to rewrite the equations of motion in the static case in a form that is suitable to extend the proof of isospectrality to modes with $L>1$ and away from the eikonal limit. Second, it would pave the way to study the dynamics of partially massless fields around Kerr-de Sitter black holes. In particular, it would be interesting to see whether the isospectrality that holds for partially massless perturbations around non-rotating backgrounds will survive also in the spinning case, and contrast it to what happens for Kerr perturbations in general relativity \cite{Detweiler:1977gy,Pani:2013ija,Pani:2013wsa,Glampedakis:2017rar,Tattersall:2018axd}. Finding a symmetry principle behind these aspects, extending  the one that  we have presented here for modes with $L=1$, would be particularly important for multiple reasons. First of all, having at our disposal other examples beyond the Chandrasekhar relation in general relativity would help us shed light on the  fundamental aspects of black holes.
In addition, a symmetry principle could be used as a powerful test of gravity and as an efficient way to discriminate among wide classes of theories.  All these interesting aspects and open questions are left for future work.

Along the way, in this paper, we gave also a systematic derivation of massive and partially massless spin-2 perturbations in S(A)dS backgrounds in arbitrary dimensions.   In particular, we were able to cast the equations of motion in the useful Schr\"odinger-like form and where all coefficients depend analytically on the frequency.  We were also able to generalize the appearance of the Gregory-Laflamme instability for massive spin-2 particles in SdS to arbitrary dimensions. 
Even if higher-dimensional S(A)dS spacetimes can not be a  realistic description of physical systems, the study of the dynamics of field perturbations in these cases can be used, by contrast, as a way to better understand fundamental properties of physical black holes.
In this spirit, we have shown in this paper that, in contrast with $4D$-SdS spacetimes, the isospectrality for partially massless spin-$2$ fields ceases to hold when $D>4$. This mirrors exactly what happens for massless particles.

More generally, the methods and results of our paper can be used as a useful tool to understand the effects of new and exotic degrees of freedom on the ringdown phase of a black hole merger.

\vspace{1cm}

\acknowledgments   We would like to thank Lam Hui, Austin Joyce and Riccardo Penco for interesting and useful discussions, and Vitor Cardoso for comments on the draft.  RAR is supported by DOE grant DE-SC0011941 and Simons Foundation Award Number 555117.  LS is also supported by Simons Foundation Award Number 555117. We thank the participants of the KITP program ``Probing Effective Theories of Gravity in Strong Fields and Cosmology" for stimulating discussions.
This research was supported in part by the National Science Foundation under Grant No. NSF PHY-1748958.

\vspace{1cm}

\appendix

\section{Partially massless spin-2 equations on 4-dimensional S(A)dS spacetime}
\label{app:pmreview}

Partially massless fields are special irreducible representations that have been shown to exist for  particles with spin greater than $1$ on  Einstein spacetimes, when their masses take very particular values \cite{Deser:1983tm,Deser:1983mm,Higuchi:1986py,Brink:2000ag,Deser:2001pe,Deser:2001us, Deser:2001wx,Deser:2001xr,Zinoviev:2001dt,Garidi:2003ys,Skvortsov:2006at, deRham:2013wv,Bernard:2017tcg}. 

In this section, we briefly review the main aspects of partially massless fields in $4$ dimensions, specializing  to SdS spacetimes. Let us start recalling first the definition of Einstein space in $4$ dimensions:
\begin{equation}
 R_{\mu\nu} = \Lambda  g_{\mu\nu} \, , 
\qquad
 R = 4 \Lambda \, ,
\label{c0}
\end{equation}
where $\Lambda$ is a constant.
In the case of SdS, the background metric takes the form
\begin{equation}
\D s^2 = -f(r)\D t^2 + \frac{\D r^2}{f(r)} + r^2 \D \Omega^2 \, ,
\qquad
f(r) \equiv 1 - \frac{r_s}{r} - \frac{\Lambda r^2}{3} \, ,
\end{equation}
where $r_s$ is the Schwarzschild radius.
Sometimes, it is convenient to rewrite the metric function $f(r)$ in $D=4$ as (see App.~\ref{app:extremallimit} for a more general discussion in arbitrary dimensions)
\begin{equation}
f(r) = \frac{\Lambda}{3r}(r-r_b)(r_c-r)(r+r_b+r_c) \, ,
\end{equation}
with 
\begin{equation}
\Lambda= \frac{3}{r_b^2+r_br_c+r_c^2} \, ,
\qquad
r_s = \frac{\Lambda}{3}r_br_c(r_b+r_c) = \frac{r_br_c(r_b+r_c)}{r_b^2+r_br_c+r_c^2} \, ,
\label{fPMD4}
\end{equation}
where $r_b$ and $r_c$ represent the black hole and de Sitter horizons, respectively. 
This way of writing the background metric makes manifest  that $f(r)\geq0$ for all values of $r$ in the range $[r_b,r_c]$.
In particular, notice that, solving Eq.~\eqref{fPMD4} for $r_b$ yields
\begin{equation}
r_b = \frac{r_c}{2} \left(\sqrt{\frac{r_c+3 r_s}{r_c-r_s}} -1 \right) \, .
\end{equation}
It is easy to show that $r_b$ is a monotonically decreasing function of $r_c$. In particular, as $r_c\rightarrow+\infty$, $r_b\rightarrow r_s$. Moreover, the two horizons coincide in the extremal limit at $r_b=r_c=\frac{3}{2}r_s$. Thus, in general,  $r_s\leq r_b \leq \frac{3}{2}r_s$ and $ \frac{3}{2}r_s\leq r_c<+\infty$. The tortoise coordinate, defined by $\D r_\star/\D r = 1/f(r)$, is
\begin{equation}
r_\star(r) = \frac{3}{\Lambda} \left[ \frac{ r_b \log\left( \frac{r}{r_b}-1  \right)}{(r_c-r_b)(2r_b+r_c)}
- \frac{ r_c \log\left( 1- \frac{r}{r_c}  \right)}{(r_c-r_b)(r_b+2r_c)}
+ \frac{(r_b+ r_c ) \log\left( 1+ \frac{r}{r_b+r_c}  \right)}{(2r_b+r_c)(r_b+2r_c)} \right] \, .
\end{equation}
In order to derive the equations describing the dynamics of partially massless fields in a SdS spacetime in $D=4$, one can start from the action \eqref{Sspin-2massive}, which we rewrite as
\begin{multline}
S = \int \D^4 x \sqrt{- g} \bigg[ - \frac{1}{2} \nabla_\lambda h_{\mu\nu}  \nabla^\lambda h^{\mu\nu} 
+  \nabla_\lambda h_{\mu\nu}  \nabla^\nu  h^{\mu\lambda} 
-  \nabla_\mu h  \nabla_\nu h^{\mu\nu} 
\\
+ \frac{1}{2} \nabla_\mu h  \nabla^\mu h
+ \frac{1}{4} R \left( h_{\mu\nu}h^{\mu\nu}- \frac{1}{2}h^2 \right)
- \frac{1}{2}m^2 \left( h_{\mu\nu}h^{\mu\nu}- h^2 \right) 
\bigg] \, ,
\label{S20}
\end{multline}
where we are keeping the mass $m$  generic for the moment.
The corresponding equations of motion are
\begin{multline}
 \square h_{\mu\nu}  -  g_{\mu\nu}  \square h
-   \nabla^\lambda  \nabla_\nu  h_{\mu\lambda} 
-   \nabla^\lambda  \nabla_\mu  h_{\nu\lambda} 
+  g_{\mu\nu}  \nabla_\rho   \nabla_\sigma h^{\rho\sigma} 
+  \nabla_\mu  \nabla_\nu h  
\\
+ \frac{1}{2} R \left( h_{\mu\nu}- \frac{1}{2} g_{\mu\nu}h \right)
- m^2 \left( h_{\mu\nu}- g_{\mu\nu}h \right)=0
\, .
\label{eom0}
\end{multline}
Using the definition of the Riemann tensor, $[ \nabla_\mu,  \nabla_\nu]X^\alpha = { R^{\alpha}}_{\beta\mu\nu}X^\beta$, and the second Bianchi identity,
\begin{equation}
\nabla_\lambda R_{\mu\nu\rho\sigma} + \nabla_\rho R_{\mu\nu\sigma\lambda} + \nabla_\sigma R_{\mu\nu\lambda\rho} = 0 \, , 
\end{equation}
together with the condition
\begin{equation}
 \nabla^\mu   R_{\mu\nu\rho\sigma} =    \nabla_\rho  R_{\nu\sigma} -  \nabla_\sigma   R_{\nu\rho} = 0\, ,
\end{equation}
where the second equality follows from being on an Einstein manifold, one can compute the divergence of the Eqs.~\eqref{eom0} to obtain
\begin{align}
&   \nabla^\mu  \square h_{\mu\nu}  -  \nabla_\nu  \square h
-  \nabla_\mu \left(    \nabla_\nu  \nabla_\lambda  h^{\mu\lambda} + { R^{\mu}}_{\,\,\,\tau\lambda\nu} h^{\tau\lambda}
+ { R^{\lambda}}_{\,\,\,\tau\lambda\nu} h^{\mu\tau} \right)
 -    \nabla_\lambda  \square  {h^\lambda}_{\nu} 
- {R^\mu}_{\,\,\, \tau\mu\lambda}  \nabla^\tau  {h^\lambda}_{\nu} 
\notag\\
&\qquad
- {R^\lambda}_{\,\,\, \tau\mu\lambda}  \nabla^\mu  {h^\tau}_{\nu} 
- R_{\nu\tau\mu\lambda}  \nabla^\mu  h^{\lambda\tau} 
+  \nabla_\rho  \nabla_{\nu}    \nabla_\sigma h^{\rho\sigma} 
+ {R^{\rho}}_{\,\,\,\lambda\nu\rho} \nabla_\sigma h^{\lambda\sigma} 
+  \nabla_\nu  \square h   
\notag\\
&\qquad
+  R_{\mu\nu}  \nabla^\mu h
+ \frac{1}{2} R \left( \nabla^\mu h_{\mu\nu}- \frac{1}{2} \nabla_\nu h \right)
- m^2 \left( \nabla^\mu h_{\mu\nu}-  \nabla_\nu h \right)
\notag\\
&= 
-  R_{\tau\nu}  \nabla_\mu       h^{\mu\tau} 
- R_{\lambda\nu} \nabla_\sigma h^{\lambda\sigma} 
+  R_{\mu\nu}  \nabla^\mu h
+ \frac{1}{2} R \left( \nabla^\mu h_{\mu\nu}- \frac{1}{2} \nabla_\nu h \right)
- m^2 \left( \nabla^\mu h_{\mu\nu}-  \nabla_\nu h \right)
\notag\\
&=  
- m^2 \left( \nabla^\mu h_{\mu\nu}-  \nabla_\nu h \right) =0 \, .
\label{eom02}
\end{align}
Assuming $m^2\neq 0$, Eq.~\eqref{eom02} results in the on-shell condition
\begin{equation}
 \nabla^\mu h_{\mu\nu}-  \nabla_\nu h = 0 \, .
\label{on1}
\end{equation}
Note that the relation \eqref{on1} contains at most first order derivatives of the spin-$2$ field. Therefore, it provides $4$ constraint equations, usually referred to as  vector constraints. These can be used to eliminate $4$ degrees of freedom out of the total $10$ components of the symmetric spin-$2$ field $h_{\mu\nu}$.
Taking another divergence on \eqref{on1}, one finds the  condition
\begin{equation}
 \nabla_\mu  \nabla_\nu  h^{\mu\nu} - \square h =0 \, ,
\label{c1}
\end{equation}
which can be combined with the trace of \eqref{eom0} to find 
\begin{equation}
( 3 m^2 -2\Lambda ) h =0
\, ,
\label{eom0trace}
\end{equation}
For generic values of the spin-$2$ mass, Eq.~\eqref{eom0trace} provides an additional scalar constraint, $h=0$, which reduces the propagating degrees of freedom from $6$ to $5$, as expected for a massive spin-$2$ field. On the other hand, if $3 m^2 -2\Lambda=0$, which corresponds to saturating the Higuchi bound \cite{Higuchi:1986py}, the combination \eqref{eom0trace} vanishes off-shell. Thus,  one loses a constraint, but gains a scalar gauge symmetry---see Eq.~\eqref{gaugetrans}---which reduces the propagating degrees of freedom from $6$ to $4$ (as opposed to a constraint, which eliminates one degree of freedom, a gauge symmetry removes two degrees of freedom). The residual modes correspond to the four components of the partially massless spin-$2$ field.


In the latter case, using the equations \eqref{on1} and \eqref{c1}, and the partially massless condition \cite{deRham:2013wv,Bernard:2017tcg}
\begin{equation}
m^2=\frac{2\Lambda}{3} \, ,
\end{equation}
the equations of motion \eqref{eom0} take  the form
\begin{equation}
\left(  \square   - m^2\right) h_{\mu\nu}  
+ 2R_{\mu\tau\nu\lambda}h^{\lambda\tau} 
- \left(  \nabla_\mu  \nabla_\nu   
+ \frac{\Lambda}{3}  g_{\mu\nu}  \right) h 
=0
\, .
\label{eom0bis}
\end{equation}
Then, the residual gauge symmetry can be  used to  eliminate one component from the partially massless tensor field $h_{\mu\nu}$ and further simplify the equations of motion. In particular, one can fix the gauge in such a way that $h_{\mu\nu}$ is traceless.\footnote{Note that this is not  the gauge that we have chosen  in the main text, though.} Then, the constraints and the equations of motion boil down to
\begin{equation}
\left(  \square   - m^2\right) h_{\mu\nu}  
+ 2R_{\mu\tau\nu\lambda}h^{\lambda\tau} 
=0
\, ,
\qquad
 \nabla^\mu h_{\mu\nu} =0 \, ,
\qquad  h = 0 \, .
\label{eom3}
\end{equation}

\section{Extremal limit of Schwarzschild de Sitter spacetime}
\label{app:extremallimit}

In this section, we show how to take the extremal limit of a SdS spacetime in arbitrary dimensions and briefly review the  expression for the quasi-normal frequencies.

Let us start noting that in SdS the position of the black hole horizon is always shifted outwards with respect to a Schwarzschild spacetime (the third term in \eqref{eq:sadsf} is always negative in dS). The function $f(r)$ has a local maximum in the interval $[r_b,r_c]$, where $r_b$ and $r_c$ denote  the black hole and cosmological horizons, respectively. The location of the maximum can be expressed as
\begin{equation}
r_0 = \left[ \frac{r_s^{D-3}(D-1)(D-2)(D-3)}{4\Lambda} \right]^{\frac{1}{D-1}}  \, ,
\end{equation}
in terms of $r_s$ and $\Lambda$.
Requiring that \eqref{eq:sadsf} admits only one solution, which amounts to imposing that the maximum of $f(r)$ satisfies $f(r_0)=0$, allows to find the extremal limit in a SdS spacetime in arbitrary dimensions: 
\begin{equation}
\bar \Lambda = \frac{ (D -3) (D -2) }{2^{\frac{D-5 }{D -3}}  (D -1)^{\frac{2}{D -3}} } r_s^{-2} \, ,
\qquad
\bar r = r_s \left(\frac{D-1}{2} \right)^{\frac{1}{D-3}} \, .
\label{extremallimit}
\end{equation}
In $D=4$, the previous expressions reduce to $\bar \Lambda_{D=4}=\frac{4}{9r_s^2}$ and  $\bar r_{D=4}= \frac{3r_s}{2}$. Calling with $\epsilon$ the relative distance between each horizon and  the maximum $r_0$, let us now expand up to quadratic order in $\epsilon$. Solving $f(r_0+\epsilon)=0$, one finds
\begin{equation}
\epsilon^2 = -2f(r_0)\left( \frac{\D^2f}{\D r^2}\bigg\vert_{r_0} \right)^{-1} =  f(r_0)\frac{D-2}{2\Lambda} \, ,
\end{equation}
where
\begin{equation}
f(r_0 ) = 1- 2^{\frac{D-5}{D-1}}r_s^{\frac{2(D-3)}{D-1}}(D-1) \left[ \frac{\Lambda}{(D-1)(D-2)(D-3)} \right]^{\frac{D-3}{D-1}} \, .
\label{fr0}
\end{equation}
Since $\Lambda$ is close  to its extremal value $\bar \Lambda$, then $f(r_0)\ll1$. The positions of the black hole and  cosmological horizons are $r_b=r_0-\epsilon$ and $r_c=r_0+\epsilon$, respectively.\footnote{Notice that, defining
\begin{equation}
\bar \delta = \sqrt{1- \frac{r_s^{2(D-3)}}{4} \frac{(D-1)^{D-1}}{(D-3)^{D-3}} \left(\frac{2\Lambda}{(D-1)(D-2)} \right)^{D-3} } \, ,
\end{equation}
Eq.~\eqref{fr0} takes on the form $f(r_0)= 1- \left( 1- \bar{\delta}^2 \right)^{\frac{1}{D-1}}$ and
\begin{equation}
\epsilon = \sqrt{\frac{D-2}{2\Lambda}}\sqrt{1- \left( 1- \bar{\delta}^2 \right)^{\frac{1}{D-1}}} \, ,
\end{equation}
in agreement with \cite{Molina:2003ff}.
}
In the extremal limit, $f(r)$ can be approximated by its Taylor expansion around $r_0$,
\begin{equation}
\begin{split}
f(r) & = f(r_0) + \frac{1}{2} \frac{\D^2f}{\D r^2}\bigg\vert_{r_0}(r-r_0)^2 + O((r-r_0)^3)
\\
&	= \frac{2\Lambda}{D-2} \left[ \epsilon^2- (r-r_0)^2  \right] + O(\epsilon^3)
\\
&	= \frac{2\Lambda}{D-2} (r-r_b)(r_c-r) + O(\epsilon^3) \, .
\end{split}
\label{frextremal}
\end{equation}
In this approximation, at the first non-trivial order in $\epsilon$, the tortoise coordinate is
\begin{equation}
r_\star(r) = \int \frac{\D r}{f(r)} = \frac{1}{2\kappa}\left[ \ln(r-r_b) - \ln(r_c-r) \right]  \, ,
\label{tortoiseextremal}
\end{equation}
where
\begin{equation}
\kappa \equiv \frac{\Lambda}{D-2} (r_c-r_b) \, .
\label{kappadef}
\end{equation}
Inverting \eqref{tortoiseextremal} and plugging back into \eqref{frextremal}, 
\begin{equation}
f(r (r_\star) ) =  \frac{\kappa (r_c-r_b)}{2\cosh^2(\kappa r_\star)} \, .
\label{fextremal2}
\end{equation}
The great advantage of \eqref{fextremal2} is that it allows to obtain an  analytic formula for the quasi-normal frequencies. 

\subsection{Quasi-normal modes in the extremal limit}

Let us postulate a Schr\"odinger-like equation of the form
\begin{equation}
\frac{\D^2}{\D r_\star^2}\Psi(r) + \left(\omega^2 - f(r) U(r) \right) \Psi(r) =0 \, ,
\qquad
\frac{\D r_\star}{\D r} = \frac{1}{f(r)} \, ,
\label{SlEqel}
\end{equation}
where $\Psi(r)$ is some generic field and $f(r) U(r)$ is the potential, such that $\lim_{r_\star\rightarrow\pm\infty} f(r) U(r) =0$. The correct boundary conditions for  QNMs,  corresponding to outgoing waves in the asymptotic regions $r_\star\rightarrow\pm\infty$, are given by
\begin{equation}
\Psi  \xrightarrow[r\rightarrow r_b]{r_\star\rightarrow-\infty} \e^{-i \omega(t+r_\star)} \, ,
\end{equation}
where $r_b$ is the location of the black hole horizon, and
\begin{equation}
\Psi  \xrightarrow[r\rightarrow r_c]{r_\star\rightarrow\infty} \e^{-i \omega(t-r_\star)} \, ,
\end{equation}
where $r_c$ is the cosmological horizon.
Let us consider now the extremal limit \eqref{extremallimit}. Retaining only the leading term in the limit $r \sim \bar r \sim r_b\sim r_c$, the Schr\"odinger-like equation can be approximated by \eqref{SlEqel}
\begin{equation}
\frac{\D^2}{\D r_\star^2}\Psi(r) + \left(\omega^2 - \frac{\kappa (r_c-r_b)U(\bar r)}{2\cosh^2(\kappa r_\star)} \right) \Psi(r) =0 \, , 
\label{SchEqEx}
\end{equation}
where $U(\bar r)$ is the value of the potential $U(r)$ at the extremal $\bar r$. The Schr\"odinger-like equation \eqref{SchEqEx} admits the following analytic expression for the quasi-normal frequencies \cite{Ferrari:1984zz,PhysRevLett.52.1361} (see also \cite{Cardoso:2003sw,LopezOrtega:2006my}):
\begin{equation}
\frac{\omega}{\kappa} = -i \left(n + \frac{1}{2}\right) + \sqrt{\frac{D-2}{2\bar \Lambda}U(\bar r) -\frac{1}{4} } \, ,
\qquad
n = 0, 1\ldots
\label{QNMextremalL}
\end{equation}
Notice that for positive potentials, $U(\bar r)>0$, the imaginary part of all the quasi-normal frequencies $\omega_n$ is negative, which indicates that the oscillations are damped. Instead, if $U(\bar r)<0$, sufficiently close to the extremal limit, there is   a non-trivial subset of frequencies, corresponding to   the lowest values of $n$, whose imaginary parts change sign, signalling the presence of an instability in the spectrum.


\section{Massive and partially massless spin-2 in 4\textit{D}: even sector equations of motion}
\label{app:pmcoeffs}

We report here the expressions for the coefficients appearing in the equations of motion \eqref{massiveeqsD4}  for massive  spin-$2$ perturbations on $4$-dimensional S(A)dS spacetimes:
\begin{align}
V_\psi  & = \frac{f}{r^3 \left((L^2+L-2) r+m^2 r^3+3 r_s\right)^2} \bigg[
9 (L^2+L-2)^2 r^2 r_s+27(L^2+L-2) r r_s^2
\notag\\
&\qquad 
	+ \left((L^2+L-2) r^5+ 6 r^4 r_s \right) \left(3 (L^2+L-4) m^2-2 \Lambda  (L^2+L-2)\right)
\notag\\
&\qquad 
	+ m^6 r^9-3 m^2 r^6 r_s (m^2-4 \Lambda )
	+m^2 r^7 (L (L+1) (2 \Lambda +3 m^2)-4 \Lambda ) +27 r_s^3	
\notag\\
&\qquad
	+r^3 \left(L (L+1) ((L-2) L (L+1) (L+3)+12)+9 r_s^2 (5 m^2-2 \Lambda )-8\right)
\bigg]
 \, ,
\\
 V_{\mathcal{K}}  & = f \left( m^2 -2 \Lambda +\frac{L^2+L+4}{r^2}-\frac{6 r_s}{r^3} \right)
  \, ,
\\
 V_{G} & = \frac{f}{3 r^3 \left(\left(L^2+L-2\right) r+m^2 r^3+3 r_s\right)^2} \bigg[
 9 r_s \left(-2 \Lambda  r^3 r_s+4 r^2-6 r r_s+3 r_s^2\right)
\notag\\
&\qquad
	+3 L^4 r^2 \left(3 m^2 r^3-r+3 r_s\right)+3 L^3 r^2 \left(6 m^2 r^3-7 r+6 r_s\right)+3 L^6 r^3+9 L^5 r^3
\notag\\
&\qquad
	+3 L r \left(m^2 r^6 (3 m^2-2 \Lambda )+6 m^2 r^3 r_s-6 m^2 r^4+4 r^2-12 r r_s+9 r_s^2\right)
\notag\\
&\qquad
	+m^4 r^6 \left(21 r_s-2 r \left(\Lambda  r^2+9\right)\right)+3 m^6 r^9+3 m^2 r^3 \left(4 \Lambda  r^3 (r-r_s)+3 r_s (5 r_s-4 r)\right)
\notag\\
&\qquad
	+ 3 L^2 r \left(3 m^4 r^6+m^2 r^3 \left(-2 \Lambda  r^3-3 r+6 r_s\right)+9 r_s (r_s-r)\right)
 \bigg]
 \, ,
\end{align}
\begin{align}
S_{\mathcal{K}}^{(\psi)}  & = 
\frac{- 2 f}{3 r^3 \left(\left(L^2+L-2\right) r+m^2 r^3+3 r_s\right)^2} \bigg[
 2 \Lambda  m^2 r^6 \left(3 (L^2+L-2) r+m^2 r^3+12 r_s\right)
\notag\\
&\quad
	-6 (L^2+L-2) r^3 \left(L^2+L+3 m^2 r^2-2\right) +18 r r_s^2 \left(3 L (L+1)+4 m^2 r^2-9\right)+81 r_s^3
\notag\\
&\quad
	+3 r^2 r_s \left(6 (L-2) (L+3) m^2 r^2+3 (L-2) (L-1) (L+2) (L+3)-m^4 r^4\right)
\bigg]
 ,
\\
S_{G}^{(\psi)}  & = 
 f \frac{(L-1) L (L+1) (L+2)  }{3  \left(\left(L^2+L-2\right) r+m^2 r^3+3 r_s\right)^2} \bigg[ 3 \left(L^2+L+1\right) r_s
\notag\\
&\qquad
- r^2 \left(2 \Lambda  (L^2+L-2) r+m^2 (6 r-9 r_s)+9 \Lambda  r_s\right)
\bigg]
  \, ,
\end{align}
\begin{align}
S_{\psi}^{(\mathcal{K})}  & = f \left(m^2 +  \frac{L^2+L-2}{r^2} + \frac{3 r_s}{r^3} \right)
 \, ,
\\
S_{G}^{(\mathcal{K})}  & = 0 
  \, ,
\end{align}
\begin{align}
S_{\psi}^{(G)}   & = - \frac{1}{2} S_{\mathcal{K}}^{(G)} 
\,  ,
\\
S_{\mathcal{K}}^{(G)} & = f\frac{8 m^2 \left(r^2 \left(2 \Lambda  \left(L^2+L-2\right) r+m^2 (6 r-9 r_s)+9 \Lambda  r_s\right)-3 \left(L^2+L+1\right) r_s\right)}{L (L+1) \left(\left(L^2+L-2\right) r+m^2 r^3+3 r_s\right)^2} \, .
\end{align}

We report here the expressions for the coefficients appearing in \eqref{PMevenL2}, which describe the dynamics of partially massless spin-$2$ fields on $4$-dimensional S(A)dS spacetimes:
\begin{align}
a_1 & = \frac{1}{3 L^2 (L+1)^2 r^2 \left(\Lambda  r^3-3 r+3 r_s\right) \left(3 \left(L^2+L-2\right) r+2 \Lambda  r^3+9 r_s\right)}
\bigg[ 16 \Lambda ^3 r^8 r_s
\notag\\
&\qquad
	+ 18 \Lambda  r^2 \left(L (L+1) (L^2+L+18)-8\right) r^2 r_s+  18 \Lambda  r^2  L^3(L+1)^3 r^3 
\notag\\
&\qquad  - 72 \Lambda  r^2  (5 L (L+1)-4) r r_s^2 - 144 \Lambda  r^2  r_s^3
	-48 \Lambda ^2 r^5 r_s \left(\left(L^2+L-2\right) r+2 r_s\right)
\notag\\
&\qquad
	-27 L (L+1) r_s \left(\left(L^2 (L+1)^2+20\right) r^2+\left(L^2+L-44\right) r r_s+24 r_s^2\right)
\bigg] \, ,
\end{align}
\begin{equation}
a_2  = \frac{3 r \left(3 \left(L^2+L+4\right) r-4 \Lambda  r^3-12 r_s\right)}{L (L+1) \left(\Lambda  r^3-3 r+3 r_s\right)^2} \, ,
\end{equation}
\begin{align}
a_3 & = \frac{1}{3 L^2 (L+1)^2 r^2 \left(\Lambda  r^3-3 r+3 r_s\right) \left(3 \left(L^2+L-2\right) r+2 \Lambda  r^3+9 r_s\right)} \bigg[ 16 \Lambda ^3 r^7 r_s
\notag \\
&\qquad
	-24 \Lambda ^2 r^4 \left(L (L+1) \left(L^2+L+1\right) r^2-4 L (L+1) r r_s+2 r r_s-2 r_s^2\right)
\notag \\
&\qquad
	-18 \Lambda  L (L+1) r^2 \left(\left(L (L+1) \left(L^2+L-4\right)-4\right) r^2+3 \left(L^2+L+8\right) r r_s-22 r_s^2\right)
\notag \\
&\qquad
	+27L^2(L+1)^2 \left((L^2+L-2) L(L+1) r^2+3 (L^2+L-4)  r r_s+15  r_s^2\right)
\bigg] \, ,
\end{align}
\begin{align}
a_4 & = \frac{4 }{9 L^2 (L+1)^2 r^2}
\bigg[ -3 \Lambda  r^2 \left(2 \left(L^2+L\right)^2 r^2+(4-3 L (L+1)) r r_s-4 r_s^2\right)
\notag\\
&\qquad
	+9 L (L+1) r_s \left(\left(L^2+L-5\right) r+6 r_s\right)+4 \Lambda ^2 r^5 r_s
\bigg]
 \, ,
\end{align}
\begin{equation}
a_5 = \frac{4 r \left(3 \left(L^2+L-2\right) r+2 \Lambda  r^3+9 r_s\right)}{L (L+1) \left(\Lambda  r^3-3 r+3 r_s\right)} \, ,
\end{equation}
\begin{align}
a_6 & = \frac{1}{9 L^2 (L+1)^2 r^2 \left(\Lambda  r^3-3 r+3 r_s\right) \left(3 \left(L^2+L-2\right) r+2 \Lambda  r^3+9 r_s\right)} \bigg[
\notag\\
&\quad
	-24 \Lambda ^3 r^7 \left(-4 \left(L^2+L-5\right) r r_s+(L-1) L (L+1) (L+2) r^2-16 r_s^2\right)
\notag \\
&\quad
	-36 \Lambda ^2 r^4 \bigg(\left(9 L (L+1) \left(L^2+L+2\right)-32\right) r^2 r_s-6 (3 L (L+1)-8) r r_s^2-16 r_s^3   
\notag\\
&\quad
	+2 L \left(L \left(L \left(L^3+3 L^2+L-3\right)+2\right)+4\right) r^3\bigg)
\notag\\
&\quad
	+54 \Lambda  r^2 \bigg(\left(L (L+1) \left(L (L+1) \left(L^2+L+18\right)+40\right)-16\right) r^2 r_s +4 (5 L (L+1)-4) r_s^3
\notag\\
&\quad
	-4 (L-1) L (L+1) (L+2) r^3-2 (L (L+1) (7 L (L+1)+36)-16) r r_s^2\bigg)
\notag\\
&\quad
	-162 L (L+1) r_s \bigg(-(L-1) (L+2) (7 L (L+1)+10) r^2
\notag\\
&\quad
	+2 (L (L+1) (5 L (L+1)-8)-22) r r_s+3 (5 L (L+1)+8) r_s^2\bigg)+64 \Lambda ^4 r^{10} r_s
\bigg] \, ,
\end{align}
\begin{align}
b_1 & = \frac{6 (L (L+1) r+r_s) \left((L-1) L (L+1) (L+2) r^2+2 (L-1) (L+2) r r_s+6 r_s^2\right)}{L^2 (L+1)^2 r^3 \left(3 \left(L^2+L-2\right) r	+2 \Lambda  r^3+9 r_s\right)}
\notag\\
&\quad
	+\frac{2 \left(L^2+L\right)^2 r^2+4 L (L+1) r r_s-4 r_s^2}{L^2 (L+1)^2 r^3}+\frac{8 \Lambda  r_s}{3 L^2 (L+1)^2}+\frac{6 r-9 r_s}{r \left(\Lambda  r^3-3 r+3 r_s\right)}\, ,
\end{align}
\begin{equation}
b_2  = \frac{3 r \left(3 \left(L^2+L-4\right) r+4 \Lambda  r^3+12 r_s\right)}{L (L+1) \left(\Lambda  r^3-3 r+3 r_s\right)^2} \, ,
\end{equation}
\begin{align}
b_3 & = -\frac{36 (L (L+1) r+2 r_s) \left(\left(L^2+L-2\right) r^2+12 r r_s-15 r_s^2\right)}{3r^4 L^2 (L+1)^2 \left(3 \left(L^2+L-2\right) r+2 \Lambda  r^3+9 r_s\right)}
\notag\\
&\quad
	+\frac{4 \left(L (L+1) \left(L^2+L+1\right) r^2-(L-1) (L+2) r r_s-8 r_s^2\right)}{r^4L^2 (L+1)^2}
\notag\\
&\quad +\frac{16 \Lambda   r_s}{3 r L^2 (L+1)^2}
	+\frac{3 \left(\left(L^2+L+4\right) r-8 r_s\right)}{r^2(\Lambda  r^3-3 r+3 r_s)}
\notag\\
&\quad 
	+\frac{18 r_s (L (L+1) r+r_s) \left(\left(L^2+L-2\right) r+6 r_s\right) \left(\left(L^2+L+4\right) r-6 r_s\right)}{r^4 L^2 (L+1)^2 \left(3 \left(L^2+L-2\right) r+2 \Lambda  r^3+9 r_s\right)^2} \, ,
\end{align}
\begin{equation}
b_4 = \frac{4 r_s \left(-9 L (L+1) \left(\left(L^2+L-5\right) r+6 r_s\right)-3 \Lambda  r^2 ((5 L (L+1)-4) r+4 r_s)-4 \Lambda ^2 r^5\right)}{L^2 (L+1)^2 r^2 \left(3 \left(L^2+L-2\right) r+2 \Lambda  r^3+9 r_s\right)^2} ,
\end{equation}
\begin{equation}
b_5 = -\frac{36 r}{L (L+1) \left(\Lambda  r^3-3 r+3 r_s\right) \left(3 \left(L^2+L-2\right) r+2 \Lambda  r^3+9 r_s\right)} \, ,
\end{equation}
\begin{align}
b_6 & = \frac{1}{L^2 (L+1)^2 r^2 \left(\Lambda  r^3-3 r+3 r_s\right) \left(3 \left(L^2+L-2\right) r+2 \Lambda  r^3+9 r_s\right)^2} \bigg[
\notag\\
&\quad
	-12 \Lambda ^2 r^4 \left(2 L (L+1) \left(L^2+L+1\right) r^2+(4-9 L (L+1)) r r_s-4 r_s^2\right)
\notag\\
&\quad
	-36 \Lambda  L (L+1) r^2 \left((L-1) (L+2) \left(L^2+L+1\right) r^2-2 \left(L^2+L-6\right) r r_s-11 r_s^2\right)
\notag\\
&\quad
	+81 L^2 (L+1)^2 r_s \left(\left(L^2+L-4\right) r+5 r_s\right)+16 \Lambda ^3 r^7 r_s
\bigg] \, .
\end{align}

\section{Quadratic actions}
\label{app:quadactions}

In this section we show how to obtain the quadratic actions for the even and odd canonically normalized degrees of freedom of spin-$2$ fields in S(A)dS spacetimes. To this end, we will introduce an auxiliary field that helps integrate out the non-dynamical variables. The procedure works in any dimensions, but, for the sake of the presentation, we will explicitly present it only in $D=4$. 

\subsection{Odd sector}

Let us start from the odd sector, focusing for the moment on the modes with $L\geq2$. With the help of an auxiliary field $\sigma$, we can rewrite  the Fierz-Pauli action \eqref{Sspin-2massive} for odd modes as
\begin{multline}
S_{\text{odd}, M=0} = \sum_{L=2}^\infty \int \D t \D r \, L(L+1) \bigg[
-\frac{1}{4} \sigma^2+ \sigma \left( a_1 \dot{h}_1 + a_2 h_0' + a_3 h_0  \right) + a_4 h_0^2 + a_5 h_1^2
\\
+ a_6 h_1 h_2'  +  a_7 h_0 \dot{h}_2 + a_8 h_2^2 + a_9 \dot{h}_2^2 + a_{10} h_2'^2
\bigg] \, ,
\label{app:auxodd}
\end{multline}
where the coefficients $a_i$, which are purely functions of $r$, are fixed by requiring that, after integrating out the auxiliary field $\sigma$ from \eqref{app:auxodd}, one recovers \eqref{Sspin-2massive} with the field $h_{\mu\nu}$  decomposed according to \eqref{hPModd4D},
\begin{subequations}
\begin{equation}
a_1 =-a_2 = 1 \, ,
\qquad 
a_3 = \frac{2}{r} \, ,
\end{equation}
\begin{equation}
a_4 = - \frac{a_5}{f^2} = \frac{-2 r f'-2 f+r^2 (m^2-2 \Lambda ) +L(L+1) }{r^2 f} \, ,
\end{equation}
\begin{equation}
a_6  = - f^2a_7  = \frac{4f^2}{r^2} a_9 = -\frac{4}{r^2}a_{10} = f (L^2+L-2) \, ,
\end{equation}
\begin{equation}
a_8 = \frac{1}{4} (L^2+L-2) \left(2 r f'+2 f-r^2 (m^2-2 \Lambda )-2\right) \, .
\end{equation}
\end{subequations}
This guarantees the equivalence between \eqref{app:auxodd} and \eqref{Sspin-2massive} for odd perturbations. The advantage of the form \eqref{app:auxodd} is that it allows to integrate out $h_0$ and $h_1$ straightforwardly, as their equations of motion are algebraic,
\begin{subequations}
\label{constrH0h1aux}
\begin{align}
h_0 & = \frac{r^2 (L^2+L-2) \dot{h}_2 - r^2 f \sigma' - 2 r f \sigma}{2 \left(-2 r f'-2 f+r^2(m^2 -2 \Lambda  )+L(L+1)\right)} \, ,
\\
h_1 & = \frac{r^2 f (2-L-L^2) h_2'   + r^2 \dot{\sigma}}{2f \left(2 r f'+2 f-r^2(m^2 -2 \Lambda  )-L(L+1)\right)} \, .
\end{align}
\end{subequations}
Notice that in \eqref{app:auxodd} we have set for simplicity the magnetic quantum number $M=0$. This will not affect the final equations as the spherical symmetry guarantees that they are independent of $M$.
Plugging \eqref{constrH0h1aux} back into the action \eqref{app:auxodd} and after simple integrations by parts, one finds a quadratic Lagrangian for $h_2$ and $\sigma$ only. Canonically normalizing the fields as follows,
\begin{subequations}
\begin{align}
\Psi_1 & \equiv r \left[ \frac{L(L+1)}{2\left( -2 r f'-2 f+r^2 (m^2-2 \Lambda )+ L(L+1) \right)} \right]^{1/2}  \sigma  \, ,
\\
\Psi_2 & \equiv \left[\frac{(L-1) L (L+1) (L+2) r^2 \left(-2 r f'-2 f+r^2 (m^2-2 \Lambda )+2\right)}{2\left(  -2 r f'-2 f+r^2 (m^2-2 \Lambda ) + L(L+1) \right)}  \right]^{1/2} h_2 \, ,
\end{align}
\end{subequations}
the quadratic action takes on the form 
\begin{tcolorbox}[colframe=white,arc=0pt,colback=greyish2] \vspace{-0.4cm}
\begin{equation}
\begin{split}
S_{\text{odd}, M=0} =  \frac{1}{2} \sum_{L=2}^\infty \int \D t \D r_\star \Bigg[
&
 \dot{\Psi}_1^2 + \dot{\Psi}_2^2 - \left( \frac{\partial\Psi_1}{\partial r_\star} \right)^2 - \left( \frac{\partial\Psi_2}{\partial r_\star} \right)^2
\\ 
&\qquad\qquad   - V_{\Psi_1} \Psi_1^2   - V_{\Psi_2} \Psi_2^2 + V_{12}\Psi_1\dot{\Psi}_2
\Bigg] \, ,
\end{split}
\label{app:quadacodd} 
\end{equation}
\end{tcolorbox}
\noindent where
\begin{subequations}
\label{app:potentialsodd}
\begin{align}
V_{\Psi_1} & = \frac{f}{3 r^3 \left(L^2+L+m^2 r^2-2\right)^2} \bigg[
\left(L^2+L-2\right) m^2 r^2 \left((9 L (L+1)-3) r+\Lambda  r^3-24 r_s\right)
\notag \\
&\quad
	+3 \left(L^2+L-2\right)^2 (L (L+1) r-3 r_s)+m^4 r^4 \left(9 L (L+1) r-2 \Lambda  r^3-24 r_s\right)+3 m^6 r^7
\bigg]
  \, ,
\\
V_{\Psi_2} & = -\frac{f}{3 r^2 \left(L^2+L+m^2 r^2-2\right)^2} \bigg[ -3 m^6 r^6
\notag \\
&\quad 
	-3 \left(L^2+L-2\right)^2 \left(L^2+L-2 \Lambda  r^2\right)
+ m^4 r^3 \left(-9 \left(L^2+L-2\right) r+2 \Lambda  r^3-3 r_s\right) 
\notag\\
&\quad +\left(L^2+L-2\right) m^2 r \left((21-9 L (L+1)) r+5 \Lambda  r^3-12 r_s\right)
\bigg]
 \, ,
\\ 
V_{12} &  = \frac{4 f m \sqrt{L^2+L-2} }{L^2+L-2+m^2 r^2} \, .
\end{align}
\end{subequations}
The result \eqref{app:quadacodd} holds for every value of the mass $m$. In particular, it  holds for partially massless fields upon substituting  $m^2 =\frac{2 \Lambda }{3}$. What is nice about this result is that it allows to   obtain equations of motion directly in a Schr\"odinger-form and  it makes manifest that they  decouple  in the  limit $m^2 =\frac{2 \Lambda }{3}\rightarrow0$. In particular, one can check that they correctly reproduce the Regge-Wheeler equation  for massless  spin-$2$ black hole perturbations  \cite{Regge:1957td} and the equation for massless spin-$1$  fields in a Schwarzschild background.  

The action for the modes with $L=1$ can be obtained following an analogous procedure, or it can more  easily be inferred from \eqref{app:quadacodd} by setting $\Psi_2\rightarrow0$ and $L\rightarrow1$:
\begin{tcolorbox}[colframe=white,arc=0pt,colback=greyish2]
\begin{equation}
S_{\text{odd}, M=0}^{(L=1)} =  \frac{1}{2}  \int \D t \D r_\star \left[
 \dot{\Psi}_1^2  - \left( \frac{\partial\Psi_1}{\partial r_\star} \right)^2    - V_{\Psi_1}^{(L=1)} \Psi_1^2   
\right] \, ,
\label{app:quadacoddL1}
\end{equation}
\end{tcolorbox}
\noindent where
\begin{equation}
V_{\Psi_1}^{(L=1)}  = f \left(  m^2 -\frac{2 \Lambda }{3} - \frac{8 r_s}{r^3}+\frac{6 }{r^2} \right) \, ,
\end{equation}
which correctly reproduces our previous result \eqref{VQodd4DL1}.

\subsection{Even sector}

Let us discuss now the even sector. Our starting point is the Lagrangian \eqref{action4DMassive}, from where one would like to integrate $\chi$ out.  In this section we will show how this can be done at the level of the action.

First, it is convenient to redefine $\mathcal{K}$ as follows,
\begin{equation}
\mathcal{K} \equiv \tilde{\mathcal{K}} +\frac{3\psi}{3 \left(L^2+L-2\right) r+2 \Lambda  r^3+9 r_s} 
+ \frac{3  L (L^2-1) (L+2) (L (L+1) r+r_s)}{12 \left(L^2+L+1\right) r_s-4 \Lambda  r^2 r_s} G .
\label{app.kcal}
\end{equation}
This allows to remove the operators $\psi'^2$ and $\psi' G$ from the Lagrangian \eqref{action4DMassive}. For the sake of the presentation, we will consider   below the case of partially massless fields. Setting $m^2 =\frac{2 \Lambda }{3}$ will indeed considerably simplify the final expressions, as we can use  the partially massless gauge freedom to  set e.g.~$\tilde{\mathcal{K}}=0$. However, one should keep in mind  that the procedure is general and it can be straightforwardly adapted for fields with generic mass.

Then, we introduce the following action 
\begin{multline}
S_{\text{even}, M=0}^{\text{PM}} = \sum_{L=2}^\infty \int \D t \D r  \bigg[
- \zeta^2+  \zeta \left( a_1 \dot{\psi} + a_2 \dot{G} + a_3 \chi' + a_4 \chi  \right) + a_5\dot G^2 +  a_6 G'^2
\\
+ a_7 G^2  +  a_8 \psi G +  a_9 \psi^2 + a_{10}\chi^2 
\bigg] \, ,
\label{app:auxeven}
\end{multline}
where $\zeta$ is an auxiliary field and where the coefficients $a_i$ are fixed in such a way that \eqref{app:auxeven} is equivalent to \eqref{action4DMassive} after plugging in  \eqref{app.kcal}  with $\tilde{\mathcal{K}}=0$ and $m^2 =\frac{2 \Lambda }{3}$. The advantage of the form \eqref{app:auxeven} is that, after simple integrations by parts, one can easily integrate out both $\chi $ and $\psi$, obtaining an action for $\zeta $ and $G$ only. Canonically normalizing the fields as
\begin{align}
\Phi_1 & \equiv   \zeta \left[ 6 \left(3 L (L+1) r \left(\left(L^2+L-2\right) r+3 r_s\right)+2 \Lambda  r^3 (2 L (L+1) r+r_s)\right)^2
\right]^{1/2}
\notag\\
&\quad
	\cdot \bigg[
12 \Lambda ^2 r^4 \left(\left(L^2+L-2\right) \left(L^2+L\right)^2 r^2-2 (2 L (L+1)+1) r_s^2\right)
\notag\\
&\quad 
	+36 \Lambda  L (L+1) r^2 \bigg((L-1) L (L+1) (L+2) \left(L^2+L-1\right) r^2
\notag\\
&\quad
	+3 (L-1) L (L+1) (L+2) r r_s+(2 L (L+1)-1) r_s^2\bigg)
\notag\\
&\quad
	+27 L^3 (L+1)^3 \left(\left(L^2+L-2\right) r+3 r_s\right)^2+8 \Lambda ^3 r^6 r_s^2
\bigg]^{-1/2} ,
\end{align}
\begin{align}
\Phi_2 & \equiv  
 G \frac{r \sqrt{L (L+1) \left(L^2+L-2\right)} }{2 \sqrt{2} [ \Lambda  r^2 r_s-3 \left(L^2+L+1\right) r_s]} \bigg[
 6 \Lambda  L^2(L^2+L-2) (L+1)^2 r^4
\notag\\
&\quad
	+9 L^2 (L+1)^2 \left((L^2+L-2) r+3 r_s\right)^2-6 \Lambda  (5 L (L+1)+2) r^2 r_s^2+4 \Lambda ^2 r^4 r_s^2
\bigg]^{1/2} ,
\end{align}
we find the following final quadratic action for the (even) partially massless degrees of freedom,
\begin{tcolorbox}[colframe=white,arc=0pt,colback=greyish2] \vspace{-0.4cm}
\begin{equation}
\begin{split}
S_{\text{even}, M=0}^{\text{PM}} =  \frac{1}{2} \sum_{L=2}^\infty \int \D t \D r_\star \Bigg[
&
 \dot{\Phi}_1^2 + \dot{\Phi}_2^2 - \left( \frac{\partial\Phi_1}{\partial r_\star} \right)^2 - \left( \frac{\partial\Phi_2}{\partial r_\star} \right)^2
\\ 
&\qquad\qquad   - V_{\Phi_1} \Phi_1^2   - V_{\Phi_2} \Phi_2^2 + V_{12}\Phi_1\dot{\Phi}_2
\Bigg] \, ,
\end{split}
\label{app:quadaceven}
\end{equation}
\end{tcolorbox}
\noindent where
\begin{align}
V_{\Phi_1} & =  \frac{1}{\left(6 (\lambda -2) \lambda ^2 \Lambda  r^5-6 (5 \lambda +2) \Lambda  r^3 r_s^2+4 \Lambda ^2 r^5 r_s^2+9 \lambda ^2 r ((\lambda -2) r+3 r_s)^2\right)^2}
\notag\\
&\quad 
\cdot	\bigg[
	-24 \Lambda ^3 r^5 r_s^2 \left(2 (\lambda -2) \lambda ^2 (\lambda +1) r^3+(\lambda -2) \lambda ^2 r^2 r_s-(\lambda -2) (6 \lambda -1) r r_s^2-8 (5 \lambda +2) r_s^3\right)
\notag\\
&\qquad +18 \Lambda ^2 r^3 \big((\lambda -2)^3 \lambda ^4 r^5+4 (\lambda -2) \lambda ^2 (7 \lambda  (3 \lambda -1)+2) r^3 r_s^2
\notag\\
&\qquad
	+4 (\lambda -2) \lambda ^2 (43 \lambda +4) r^2 r_s^3+14 (\lambda -2)^2 \lambda ^4 r^4 r_s+\lambda  (\lambda  (179 \lambda -14)+32) r r_s^4
\notag\\
&\qquad
	-6 (\lambda  (13 \lambda +20)+4) r_s^5\big)
-54 \lambda ^2 \Lambda  r^2 \big((\lambda -2)^3 \lambda ^2 r^4+2 (\lambda -2) \lambda  (\lambda  (29 \lambda -8)-10) r^2 r_s^2
\notag\\
&\qquad
	+8 (\lambda -2)^2 \lambda ^2 (\lambda +1) r^3 r_s+12 (\lambda -2) (3 \lambda +1) (4 \lambda +1) r r_s^3+9 (\lambda  (18 \lambda +11)+2) r_s^4\big)
\notag\\
&\qquad
	-8 \Lambda ^4 r^7 r_s^4 ((5 \lambda +2) r+16 r_s)+81 \lambda ^5 ((\lambda -2) r+3 r_s)^4
\bigg] \, ,
\end{align}
\begin{align}
V_{\Phi_2} & =  \frac{1}{r^3 \left(6 (\lambda -2) \lambda ^2 \Lambda  r^4-6 (5 \lambda +2) \Lambda  r^2 r_s^2+4 \Lambda ^2 r^4 r_s^2+9 \lambda ^2 ((\lambda -2) r+3 r_s)^2\right)^2}
\notag\\
&\quad
	\cdot \bigg[
	24 \Lambda ^3 r^6 r_s^2 \left(2 (\lambda -2) \lambda ^2 (2 \lambda -1) r^3+17 (\lambda -2) \lambda ^2 r^2 r_s+(\lambda -2) (9 \lambda -1) r r_s^2-8 (\lambda +1) r_s^3\right)
\notag\\
&\qquad
	+18 \lambda  \Lambda ^2 r^4 \big((\lambda -2)^2 \lambda ^3 (7 \lambda -2) r^5 +4 (\lambda -2) \lambda  (\lambda  (24 \lambda -13)+2) r^3 r_s^2
\notag\\
&\qquad
	+4 (\lambda -2) \lambda  (31 \lambda -20) r^2 r_s^3+32 (\lambda -2)^2 \lambda ^3 r^4 r_s+(\lambda -2) (161 \lambda -16) r r_s^4
\notag\\
&\qquad +12 (11 \lambda -4) r_s^5\big)
	+54 \lambda ^2 \Lambda  r^2 \big((\lambda -2)^3 \lambda ^2 (3 \lambda -1) r^5+(\lambda -2) \lambda  (\lambda  (41 \lambda -146)+20) r^3 r_s^2
\notag\\
&\qquad	
	+6 (\lambda -2) (\lambda  (11 \lambda -8)+2) r^2 r_s^3+16 (\lambda -2)^3 \lambda ^2 r^4 r_s-9 (\lambda -2) (2 \lambda +5) r r_s^4
\notag\\
&\qquad
	-72 (\lambda +1) r_s^5\big) +8 \Lambda ^4 r^8 r_s^4 ((\lambda -2) r+2 r_s)
\notag\\
&\qquad
	+81 \lambda ^4 ((\lambda -2) r+3 r_s)^2 \left((\lambda -2)^2 \lambda  r^3+3 (\lambda -2)^2 r^2 r_s+9 (\lambda -2) r r_s^2+9 r_s^3\right)
\bigg] \, ,
\end{align}
\begin{align}
V_{12} & =  -\frac{324 \Lambda^{3/2}r^{6} r_s \left(3 \lambda -\Lambda  r^2+3\right) \left(4 \lambda  \Lambda  r^3+2 \Lambda  r^2 r_s+3 (\lambda -2) \lambda  r+9 \lambda  r_s\right)}{\sqrt{\lambda -2} \lambda  \left(3 \lambda +2 \Lambda  r^2\right) \left(\Lambda  r^3-3 r+3 r_s\right)^3 \left(2 \Lambda  r^3+3 (\lambda -2) r+9 r_s\right)^4}
\notag\\
&\quad \cdot \bigg[
	(\lambda -2) \left(\sqrt{4 \lambda +1}-1\right) \left(\sqrt{4 \lambda +1}+1\right) \left(3 \lambda +2 \Lambda  r^2\right) \left(\Lambda  r^3-3 r+3 r_s\right)^3 
\notag\\
&\qquad
	\cdot\left(2 \Lambda  r^3+3 (\lambda -2) r+9 r_s\right)^4 \left(3 (\lambda -2) \lambda ^2 r^2+r_s^2 \left(9 \lambda +2 \Lambda  r^2\right)+18 \lambda ^2 r r_s\right)
\bigg] 
\notag \\
&\quad 
	\cdot\bigg[ 54 \sqrt{6} \Lambda  r^6 r_s \left(-3 \lambda +\Lambda  r^2-3\right) \left(4 \lambda  \Lambda  r^3+2 \Lambda  r^2 r_s+3 (\lambda -2) \lambda  r+9 \lambda  r_s\right)
\notag \\
&\qquad 
	\cdot \bigg(2 \Lambda  r^4 \left(3 \lambda ^3-6 \lambda ^2+2 \Lambda  r_s^2\right)+3 r^2 \left(3 \lambda ^4-12 \lambda ^3+12 \lambda ^2-10 \lambda  \Lambda  r_s^2-4 \Lambda  r_s^2\right)
\notag \\
&\qquad +54 (\lambda -2) \lambda ^2 r r_s+81 \lambda ^2 r_s^2\bigg)
	\bigg]^{-1}
\, ,
\end{align}
where we defined $\lambda\equiv L(L+1)$. Notice that, sending $\Lambda\rightarrow0$, the equations of motion  for $\Phi_1$ and $\Phi_2$, obtained from \eqref{app:quadaceven}, decouple  and one correctly recovers the Zerilli equation \cite{Zerilli:1971wd} for massless spin-$2$ perturbations and the equation of motion for massless spin-$1$ fields on Schwarzschild backgrounds.

So far, we have considered the case of modes with generic $L$. The particular case of $L=1$ can be easily inferred from the previous expressions by simply setting $G\rightarrow0$ ($\Phi_2\rightarrow0$) and $L\rightarrow1$. The canonically-normalized quadratic action takes on the form,
\begin{tcolorbox}[colframe=white,arc=0pt,colback=greyish2] 
\begin{equation}
S_{\text{even}, M=0}^{\text{PM} \, (L=1)} =  \frac{1}{2}  \int \D t \D r_\star \left[
 \dot{\Phi}_1^2  - \left( \frac{\partial\Phi_1}{\partial r_\star}  \right)^2
  - V_{\Phi_1}^{(L=1)} \Phi_1^2  
\right] \, ,
\label{app:quadacevenL1}
\end{equation}
\end{tcolorbox}
\noindent where
\begin{equation}
\Phi_1 \equiv \frac{r_s \sqrt{\frac{\Lambda  r^2}{3}+1} \left(\Lambda  r^2-9\right)}{\Lambda  r^3 (4 r+r_s)+9 r r_s} \zeta \, .
\end{equation}
and where the potential is
\begin{equation}
 V_{\Phi_1}^{(L=1)}
= 2 f\frac{\Lambda  r^2 (\Lambda  r (3 r+4 r_s)+54)-81}{r^2 \left(\Lambda  r^2-9\right)^2} \, ,
\end{equation}
in agreement with \eqref{potevenL1PM}.

\bibliographystyle{JHEP}
\addcontentsline{toc}{section}{References}
\bibliography{biblio}

\end{document}